\documentclass[a4paper,fleqn,usenatbib,useAMS]{mnras}

\usepackage{graphicx,multicol} 
\usepackage{times}
\def\etal{\it et~al.}

\title[Emission in different modes of PSR J0826+2637]{Radio emission features in different modes of PSR J0826+2637 (B0823+26)}
\author[Basu \& Mitra]{Rahul Basu$^{1,2}$, Dipanjan Mitra$^{3,2}$ \\
$^{1}$ Inter-University Centre for Astronomy and Astrophysics, Pune, 411007, India; rahulbasu.astro@gmail.com \\
$^{2}$ Janusz Gil Institute of Astronomy, University of Zielona G\'ora, ul. Szafrana 2, 65-516 Zielona G\'ora, Poland \\
$^{3}$ National Centre for Radio Astrophysics, Tata Institute of Fundamental Research, Pune 411007, India \\
}
\begin{document}



\maketitle

\label{firstpage}

\begin{abstract}
We report on a detailed analysis of the radio emission during the different 
modes of the pulsar J0826+2637 (B0823+26), observed using the Giant Meterwave 
Radio Telescope at 306-339 MHz observing frequencies. The pulsar profile has a 
postcursor and interpulse emission in addition to the main pulse. The single 
pulses showed the presence of nulling, periodic fluctuation in the emission as 
well as two prominent modes. In addition the pulsar also showed the presence of
a null state where no emission was seen for roughly an hour which was 
immediately followed by a short duration ($\sim$5 minutes) bright state termed 
the Q-bright state. The nulling varied significantly in the two modes, from a 
few percent nulls in B-mode to more than 90 percent nulling during the Q-mode. 
Additionally, the pulsar showed the presence of low level emission in both the 
interpulse and postcursor components when the main pulse nulled in B-mode. We 
detected periodic fluctuations in both the main pulse and postcursor during 
B-mode which were most likely a form of periodic amplitude modulation unrelated
to subpulse drifting. We have also detected the appearance of periodicity 
during the transitions from the null to the burst states in the Q-mode, which 
was longer than the B-mode modulations. Our analysis further revealed a 
significant increase in the main pulse and post-cursor intensity during the 
transition from the Q-mode to the short duration Q-bright mode. On the other 
hand no commensurate variation was visible in the interpulse intensity. 
\end{abstract}


\section{Introduction}
\noindent
In rotation powered pulsars the coherent radio emission is typically a tiny 
fraction ($<$1\%) of the rotational energy, and is thought to arise due to 
growth of instabilities in relativistically flowing pair-plasma. When several 
thousand individual pulses are averaged a stable pulse profile is formed. The 
shape of the profile appears to remain identical across assorted observations 
at various observing times. However, a small subsample ($\leq$ 30 pulsars) 
shows transitions between more than one stable state with distinct emission 
properties and profile shapes. This phenomenon is known as emission mode 
changing, also sometimes called state change \citep{bac70a,bar82,wan07}. There 
is also the phenomenon of nulling where the pulsar switches to a state where 
the radio emission disappears completely, or at least goes below detection 
limits \citep{bac70b,big92,wan07,gaj12,bas17}. The durations of the modes last 
between several minutes to hours at a time, and the transition between them is 
very rapid, usually within one rotation period. One of the prominent examples 
of mode changing is seen in the pulsar J0826+2637 (B0823+26), where at least 
two distinct modes are present characterised by their `Quiet' and `Bright' 
emission states lasting for hours at a time \citep{sob15}. In a recent study by
\cite{her18} the radio mode changing in this pulsar is also seen to be 
accompanied by synchronous variation in the X-ray emission. The X-ray emission 
is incoherent in nature and carries a larger fraction of the rotational energy 
of the pulsar compared to the coherent radio emission. Hence, the synchronous 
X-ray and radio mode changes suggest that the physical mechanism responsible 
for this phenomenon affects both the coherent and incoherent emission 
processes. There are also a few pulsars which exhibit intermittent behaviour 
where they spend large amounts of time (months to years) in the radio active 
and null states, respectively \citep{kra06,lyn10}. The state changes in these 
systems were accompanied by changes in their rotational energies. A detailed 
timing analysis was carried out for PSR J0826+2637 by \citet{you12}, but the 
authors failed to uncover any detectable changes in the spin-down rate. There 
are also examples of quasi-periodic variations in intensity in the form of 
periodic amplitude modulation and periodic nulling. In a series of studies 
\citet{bas16,bas17,mit17,bas18b,bas19a,bas19b} have identified and delineated 
the properties of these phenomena whose physical properties were clearly 
different from subpulse drifting.\footnote{The phenomenon of subpulse drifting 
is associated with plasma columns performing ${\bf E} \times {\bf B}$ drift at 
the surface of the polar cap and is identified as a local phenomenon \citep[see
e.g.][]{rud75}, unrelated to mode changing.}

In order to explain mode changes, one needs to appeal to a mechanism
which causes the radio emission to change rapidly, and subsequently exhibit 
stable radio emission with different modal characteristics. The mode changes 
are most likely associated with sudden changes in the properties of the radio 
emitting relativistic plasma as suggested by \cite{bar82,bas18b,bri19}. 
However, no physical mechanism has been identified that can explain :\\
(a) what triggers the fast change in the plasma flow;\\
(b) what decides the stability of the plasma flow when the pulsar resides in a 
certain mode;\\
(c) and what causes it to switch back and forth between the different modes.\\ 
In the literature there are a few suggestions regarding processes that can
change the plasma flow resulting in mode changing. The changes in conditions of
the partially screened gap \citep[PSG,][]{gil03,sza15} in the inner 
acceleration region, or variations in the non-dipolar magnetic fields above the
polar cap due to Hall-drift \citep[][Geppert et al. in preparation]{gep14}
are examples of local changes occurring above the polar cap. Alternatively, 
changes in viewing geometry that arise due changes of size of the global 
magnetosphere \citep{tim10}, or transition from a non-corotating to co-rotating
magnetosphere \citep{yue17} are other examples of such variations in the larger 
magnetosphere. Thus intensive studies of the emission properties in known 
objects like PSR J0826+2637 as well as uncovering other diverse examples of 
these phenomena will be important to test the viability of the different models
explaining the physical processes in the pulsar magnetosphere.

The average profile of PSR J0826+2637 shows the presence of a post-cursor 
component and an inter-pulse along with the main pulse \citep{bac73}. The 
inter-pulse is roughly separated by 180\degr~in longitude from the main pulse 
indicating a nearly orthogonal emission geometry \citep{han86}. The main pulse
shows the presence of lower linear polarization as well as sign changing 
circular polarization which are indicative of a core component. The 
polarization position angle (PPA) shows complicated behaviour due to the 
presence of orthogonal polarization modes. \cite{ran95} have carried out 
detailed polarization mode separated studies to estimate around 120\degr~swing 
of the PPA across the main pulse. This is consistent with a central line of 
sight traverse of the emission beam as suggested by the rotating vector model 
\citep{rad69}. This clearly suggests that the main pulse corresponds to a core
single (S$_t$) profile. The post-cursor has significant evolution and 
apparently its separation from the main pulse changes as a function of 
frequency. However, careful studies by \citet{bas15} revealed the outer edge of
the post-cursor component to be frequency invariant, thereby indicating a 
constant separation from the main pulse. The pulsar also shows large diversity 
in its single pulse behaviour. In addition to the mode changing mentioned 
earlier the pulsar exhibits the presence of nulling at both long and short 
intervals \citep{you12}. The single pulses fluctuate periodically which are 
likely to be manifestation of the periodic amplitude modulation \citep{wel06,
wel07b,sob15}. \citet{mit15} have carried out careful analysis of the stronger 
single pulses from this source and identified the presence of micro-structures 
in the main pulse. The micro-structures were interpreted as signatures of 
temporal modulations in the plasma flow responsible for the radio emission.

In the recent study reported by \citet{her18} a comprehensive observing 
campaign was undertaken to monitor the pulsar J0826+2637~at X-ray frequencies
using \textit{XMM-Newton}, which was shadowed at radio frequencies primarily by
the Giant Meterwave Radio Telescope (GMRT) and supported with LOw Frequency 
ARray (LOFAR) international stations. The primary goal of the radio 
observations in this work was to identify the intervals of the different modes 
and determine the commensurate X-ray variations. In the process the GMRT 
performed highly sensitive observations of a large number of single pulses in 
different emission states. In this paper we have concentrated on understanding 
the detailed nature of radio emission observed using GMRT during the different 
emission modes. We have analysed the single pulses to characterise the 
different phenomena like nulling, periodic fluctuations, etc., in the different
modes. In section \ref{sec:obs} we have summarized the observing details and 
the principal modal behaviour discussed in the previous work. Section 
\ref{sec:null} reports the nulling properties during each observing session 
while section \ref{sec:ampmod} details the periodic modulations in the single 
pulse behaviour. In section \ref{sec:post} we investigate the bursts seen in 
the post-cursor component particularly at the start of the B-mode as well as 
the inter-pulse. We have carried out a detailed discussion (section 
\ref{sec:disc}) of the physical implications of the single pulse behaviour seen
in the different emission modes. Finally, we summarize the results of our 
studies in section \ref{sec:sum}.

\section{Observation and Emission states}\label{sec:obs}

\begin{table}
\resizebox{\hsize}{!}{
\begin{minipage}{80mm}
\caption{Observations of PSR J0826+2637 at 339 MHz using GMRT}
\centering
\begin{tabular}{cccc}
\hline
 Date & Duration & Pulses & Mode \\
  & (hrs) &  &  \\
\hline
 20 April, 2017 & 8.1 & 41470 & B \\
 22 April, 2017 & 5.1 & 32046 & B \\
 24 April, 2017 & 8.3 & 32181 & Q, Q-Bright \\
 26 April, 2017 & 4.0 & 26001 & B \\
 28 April, 2017 & 6.7 & 30637 & B \\
 30 April, 2017 & 3.9 & 24656 & B \\
\hline
\end{tabular}
\label{tabobs}
\end{minipage}
}
\end{table}

\begin{figure*}
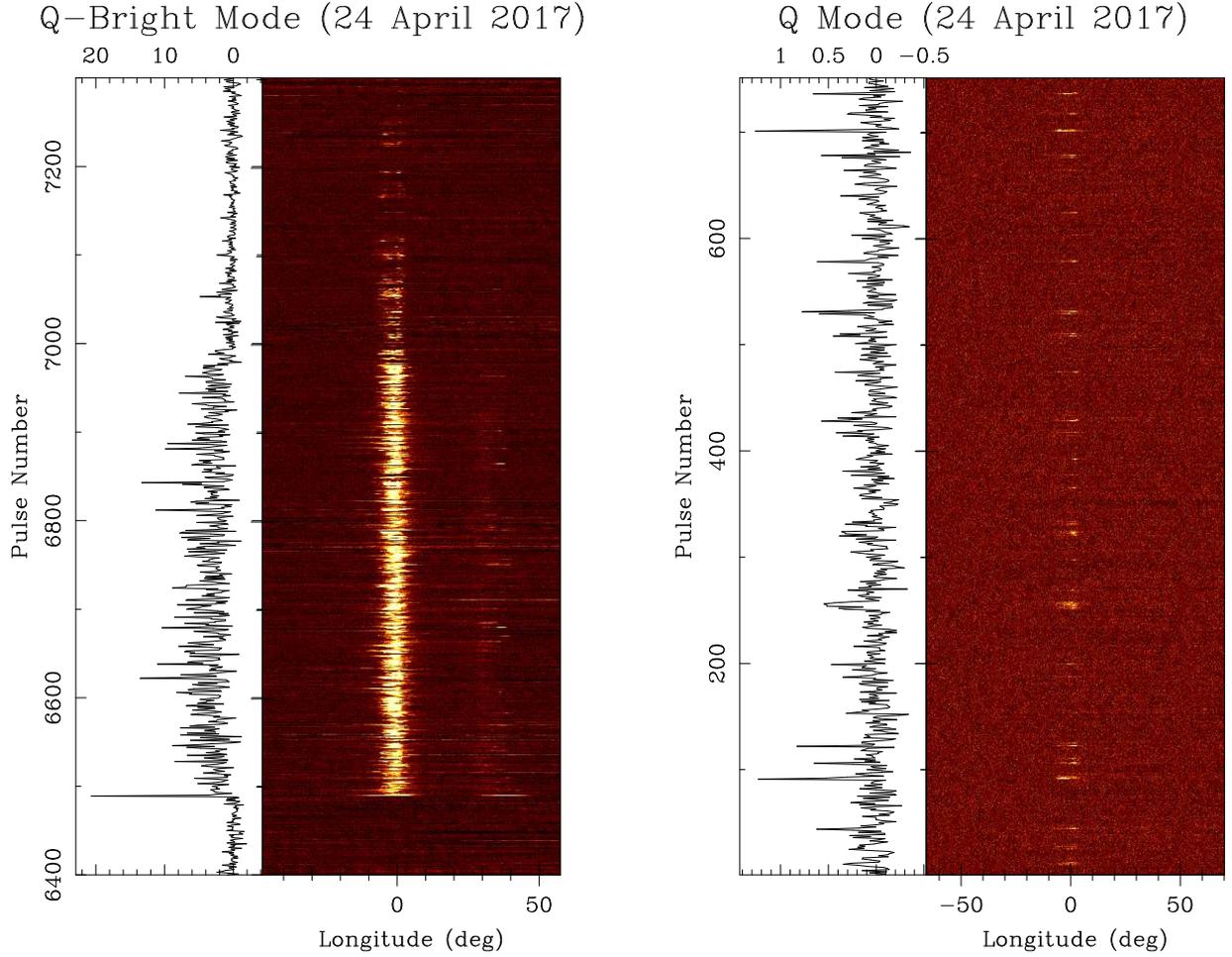

\begin{tabular}{@{}cr@{}}
{\mbox{\includegraphics[scale=0.5,angle=0.]{J0826_Q-bright.ps}}} &
\hspace{20px}
{\mbox{\includegraphics[scale=0.5,angle=0.]{J0826_Q.ps}}} \\
\hspace{30px}
\end{tabular}
\caption{The figure shows the emission from PSR J0826+2637 during 24 April 2017
when the pulsar was primarily in the Q-mode. The right window in each panel 
shows the intensity variations of the single pulses (in the colour scale) and 
the left window shows the average intensity of each pulse (in arbitrary 
counts). The left panel shows the transition from the null state to the short 
duration Q-bright state. The unusually bright pulse at the start of this state 
is clearly seen in the figure. The right panel shows the typical Q-mode of the 
pulsar which is mainly dominated by nulls with low level emission seen for 
short durations.}
\label{fig_modesngl}
\end{figure*}

\begin{figure}
\begin{center}
\includegraphics[scale=0.75,angle=0.]{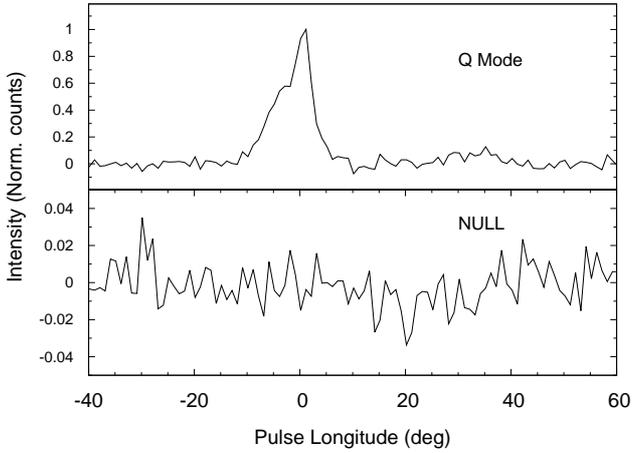}
\end{center}
\caption{The figure shows the folded profile of PSR J0826+2637 during different
emission states on 24 April, 2017. The top panel corresponds to the Q-mode 
where the emission is much weaker. The pulsar profile is wider and peaks near 
the trailing edge and possibly exhibits a weaker double component. The 
post-cursor component is not clearly seen in the average profile. The bottom 
panel corresponds to the nulling interval which lasts for around 5500 periods 
before the start of the short duration bright state. The folded profile shows 
the absence of any low level emission during this state.}
\label{fig_Qfold}
\end{figure}

\begin{figure}
\begin{center}
\includegraphics[scale=0.75,angle=0.]{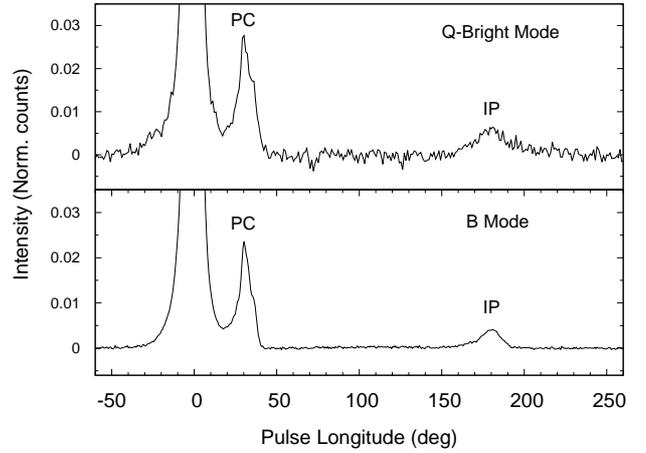}
\end{center}
\caption{The figure shows the folded profile of PSR J0826+2637 during the short
duration bright state lasting around 500 periods during the Q-mode (top panel),
and a typical example of B-mode observed on 26 April, 2017. In the absence of 
absolute flux scaling the peak of the main pulse has been normalized to unity.
We have highlighted the post-cursor component and interpulse in the figure. 
Both the post-cursor and interpulse appear to be stronger compared to the main 
pulse during the Q-Bright state. However, this may also be an outcome of the 
smaller number of pulses used in this profile resulting in insufficient 
sampling.}
\label{fig_Bfold}
\end{figure}

\begin{table*}
\caption{Average Component Properties in the different Modes}
\centering
\begin{tabular}{cc@{\hskip3pt}cc@{\hskip3pt}cc@{\hskip3pt}c@{\hskip3pt}cc@{\hskip3pt}cc}
\hline
 MODE & \multicolumn{2}{c}{Main Pulse} & \multicolumn{2}{c}{Postcursor} & \multicolumn{3}{c}{Inter Pulse} & \multicolumn{2}{c}{PC-MP} & IP-MP \\
   & W$_{50}$ & W$_{10}$ & W$_{50}$ & PC/MP & W$_{50}$ & W$_{10}$ & IP/MP & W$_{SEP}^{Peak}$ & W$_{SEP}^{Edge}$ & W$_{SEP}^{Peak}$ \\
   & (\degr) & (\degr) & (\degr) & (\%) & (\degr) & (\degr) & (\%) & (\degr) & (\degr) & (\degr) \\
\hline
   &  &  &  &  &  &  &  &  &  &  \\
 B & 4.15$\pm$0.33 & 9.94$\pm$0.33 & 7.45$\pm$0.33 & 2.36$\pm$0.01 & 13.82$\pm$0.33 & 31.85$\pm$0.33 & 0.41$\pm$0.01 & 30.70$\pm$0.33 & 40.72$\pm$0.33 & 178.83$\pm$0.33 \\
   &  &  &  &  &  &  &  &  &  &  \\
 Q-Bright & 4.52$\pm$0.33 & 10.37$\pm$0.33 & 10.75$\pm$0.33 & 2.77$\pm$0.11 & --- & --- & 0.64$\pm$0.11 & 31.03$\pm$0.33 & 40.57$\pm$0.33 & 178.17$\pm$0.33 \\
   &  &  &  &  &  &  &  &  &  &  \\
 Q & 6.77$\pm$0.33 & 14.75$\pm$0.67 & --- & --- & --- & --- & --- & --- & --- & --- \\
   &  &  &  &  &  &  &  &  &  &  \\
\hline
\end{tabular}
\label{tabcomp}
\end{table*}

\noindent
We observed total intensity single pulses from the pulsar using GMRT on six 
separate occasions in April, 2017, as listed in Table \ref{tabobs}. The GMRT 
consists of 30 antennas distributed in a Y-shaped array, with 14 antennas 
located within a central square kilometer area and the remaining 16 spread 
along three arms \citep{swa91}. The observations were carried out in the 
306-339 MHz frequency band, with a time resolution of approximately 0.5 
milliseconds. For these observations we used around 20 antennas, the available 
central square antennas and the first 2 arm antennas, whose recorded signals 
were co-added in the `Phased-Array' mode to improve detection sensitivity of 
the single pulses. The antenna phases were aligned by observing a strong nearby
point like source, and the process was repeated every one and half hours (or 
roughly ten thousand pulses for the period of 0.53 seconds). The phasing 
durations lasted between 5 to 15 minutes at a time, during which the pulsar 
emission could not observed. One of the drawbacks of these studies was the lack
of any absolute flux calibration. This required interferometric observations 
with suitable phase and flux calibrators observed at regular intervals to scale
the observed counts accordingly. Though the GMRT is capable of simultaneous 
observations in both `Interferometric' and `Phased-Array' modes, this would 
have required additional down time for calibration setups. The option of flux 
calibration was forfeited in order to maximize the duration of synchronous 
observations with \textit{XMM-Newton}. Additionally, the telescope performance 
was also reset after every phasing interval, implying the the single pulse 
intensities could not be directly compared for different intervals. However, 
the average profile of this pulsar is remarkably constant over time and at 
our observing frequencies scintillation effects were negligible \citep{das13,
sob15}. As a result we could compare the single pulse properties across phasing
intervals after normalizing using the profile peak for that duration. This was 
particularly applicable for the B-mode which was prevalent during these 
observations. We were allotted approximately 8-9 hours of observing time on 
each day with the exception of 26 April when only 5 hours of observations were 
scheduled. The initial telescope setup took between 30 minutes to 1 hour. On 
two sessions, 22 and 30 April, the observations had to be truncated by 2-3 
hours due to recording malfunction and high winds. The presence of radio 
frequency interference (RFI) during certain intervals affected our ability to 
observe the pulsar. In Table \ref{tabobs} we have listed the effective single 
pulses observed during each observing session, excluding RFI affected pulses 
and phasing intervals, which amounts to roughly 187,000 pulses over the entire 
observations.

As detailed in \citet{her18}, the pulsar was primarily seen in the B-mode 
during most of the observing run. The only exception was 24 April when the 
pulsar was in the Q-mode. A short duration bright emission state was seen on 
this day starting from pulse number 6490 and lasting for roughly 500 periods as
shown in figure \ref{fig_modesngl} (left panel). The emission gradually 
transitioned to the typical Q-mode emission, which was characterised by short 
duration low level emission interspersed between nulls (see figure 
\ref{fig_modesngl}, right panel). Before the start of the bright phase the 
pulsar was in the null state starting around pulse number 982 and lasting for 
around 5500 pulses (see figure \ref{fig_modesngl}, left panel, the initial 
pulses). \citet{her18} classified this short duration burst as a Q-bright state
to distinguish from the B-mode which lasts for hours at a time. The modal 
behaviour of the pulsar during these observations were more protracted compared
to previous studies \citep{sob15}. During the entire observing run we were not 
able to observe any clear transition from Q-mode to B-mode or vice versa. The
average profiles in the Q-mode and the nulling preceding the bright state in 
Q-mode is shown in figure \ref{fig_Qfold}. The postcursor and interpulse 
emission were not seen in the average profile of Q-mode, which was likely due 
to the emission being much weaker as well as being dominated by null pulses. 
The profile corresponding to the long duration null state did not reveal the 
presence of any low level emission. In figure \ref{fig_Bfold} the profiles of 
the bursting state in Q-mode and the more conventional B-mode are shown, where 
we have highlighted the postcursor and interpulse component. In Table 
\ref{tabcomp} we have estimated the widths of the different components and 
their separations during the different emission states, including the widths 
for the main pulse at 50\% (W$_{50}$) and 10\% (W$_{10}$) level of peak 
intensity in all three modes. The postcursor width was only estimated at 50\% 
level since the bridge emission was above the 10\% level of the postcursor 
peak. On the other hand the interpulse was only prominent in the B-mode for 
estimating the widths. The main pulse showed a single component feature during 
both the Q-Bright and B-mode, with the former having a sightly larger width. 
However, the main pulse showed considerable evolution during the Q-mode with 
the width becoming almost 50\% wider and resembling a barely resolved double 
shape. The postcursor component also became wider during the Q-Bright state 
compared to the B-mode. The Table also shows the ratio of the peak values 
of the main pulse compared to the postcursor (PC/MP) as well as the interpulse 
(IP/MP) expressed as a percentage of the main pulse peak. The ratios were 
calculated for the B-mode and the Q-bright mode profiles. The error bars were
estimated from the 1-$\sigma$ variations of the baseline for the different 
profiles and is larger for the Q-bright due to the relatively small number of 
pulses used in the profile ($\sim$ 500). Both the postcursor and interpulse are
brighter in the Q-bright mode compared to the B-mode. The single pulse emission
showed large fluctuations (see section \ref{sec:ampmod}) and the variations 
seen in the profile shape and intensities of different components between the 
Q-bright mode and the B-mode could also result from the small number of pulses 
available to form the former profile. We have also estimated the separation 
between the postcursor component and the main pulse peak at two realizations, 
between the two peaks and from the main pulse peak to the trailing edge of the 
postcursor component \citep[see discussion in][]{bas15}, which were similar in 
the two modes and consistent with previous estimates. The separations between 
the main pulse and interpulse peaks were around 178\degr~for both the B-mode 
and Q-Bright mode which were also similar to earlier measurements \citep{han86,
sob15}, and suggests the rotation and magnetic axes to be nearly perpendicular 
to each other.

\begin{figure}
\begin{center}
\includegraphics[scale=0.65,angle=-90.]{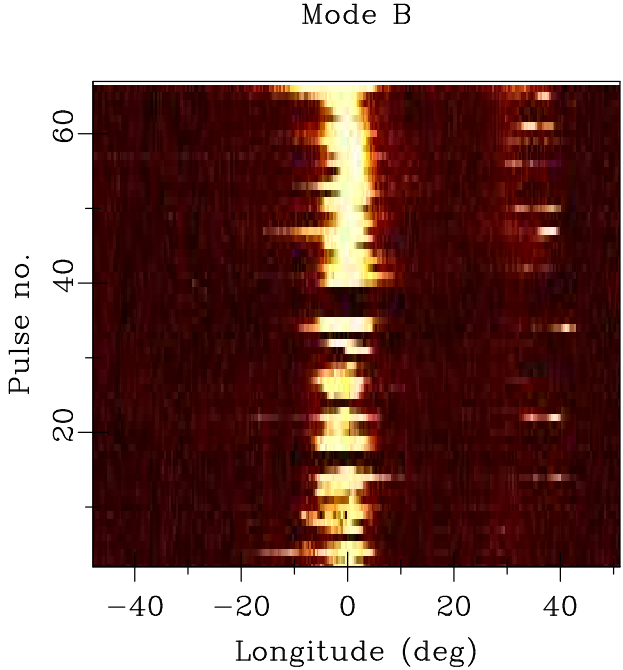}
\end{center}
\caption{The figure shows single pulses during the emission B-mode observed on 
26 April, 2017. The sequence highlights the presence of nulls seen during this 
mode, which are typically short duration lasting a few periods.}
\label{fig_Bnull}
\end{figure}

\section{Nulling}\label{sec:null}

\begin{figure}
\begin{center}
\includegraphics[scale=0.67,angle=0.]{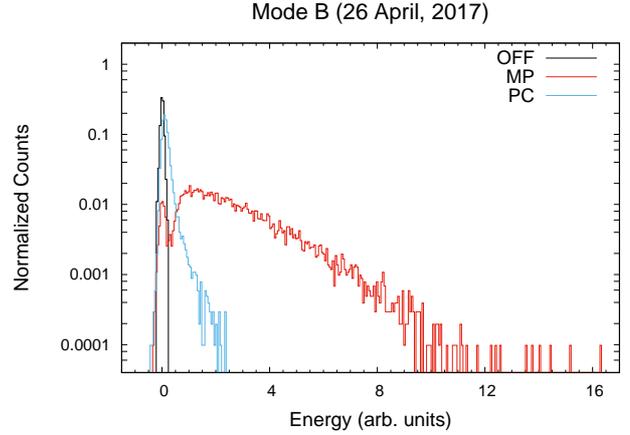}
\end{center}
\caption{The figure shows the energy distributions of the main pulse, 
postcursor and off pulse region of PSR J0826+2637~in the B-mode, observed on 26
April, 2017. The main pulse distribution shows the presence of two distinct 
regions, the null pulses showing a Gaussian distribution coincident with the 
off pulse distribution, and a long tail corresponding to burst pulses. The post
cursor on the other hand is much weaker and do not show any such distinctions.}
\label{fig_Bdist}
\end{figure}

\begin{figure*}
\begin{tabular}{@{}lr@{}}
{\mbox{\includegraphics[scale=0.68,angle=0.]{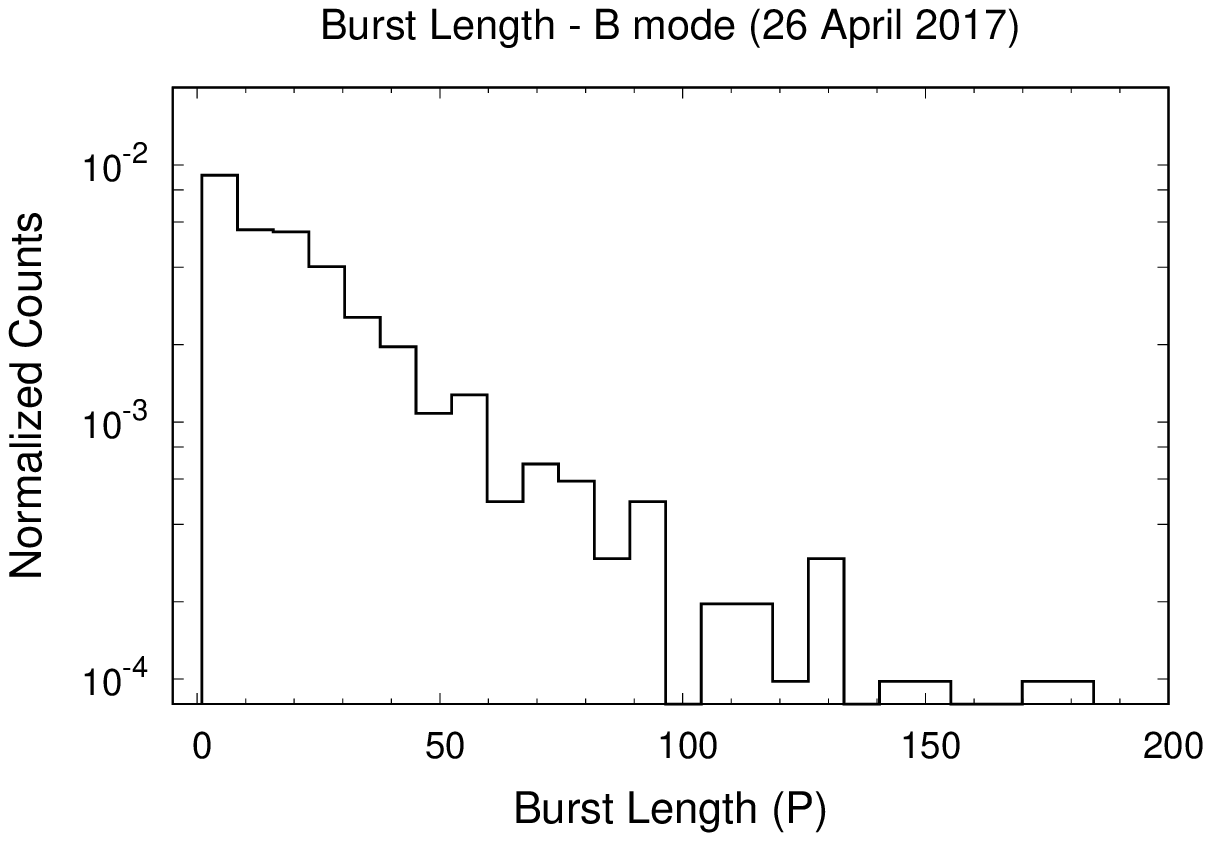}}} &
{\mbox{\includegraphics[scale=0.68,angle=0.]{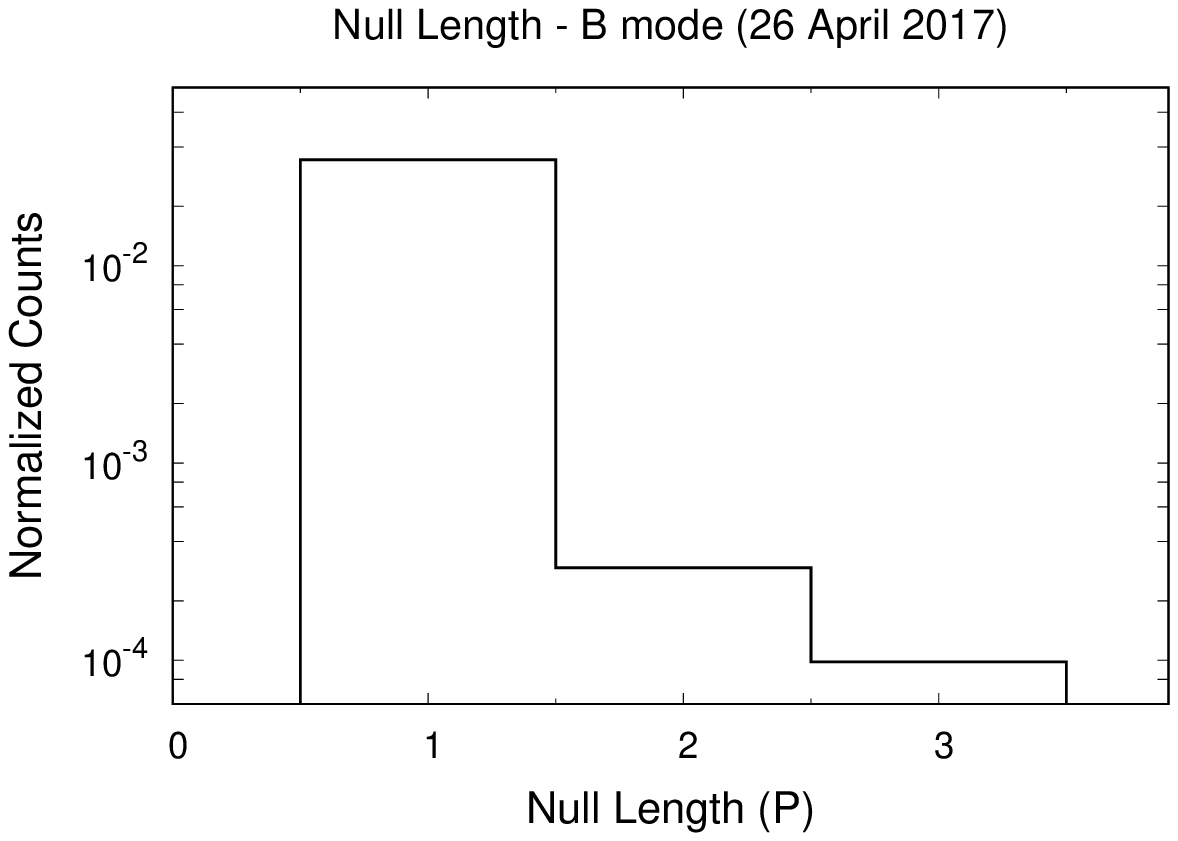}}} \\
\end{tabular}
\caption{The burst length (left panel) and null length (right panel) histograms
during B-mode, observed on 26 April, 2017. The nulls are short duration during 
this mode with the majority being a single period null. The bursting states on 
the other hand lasts for longer durations with maximum of around 200 periods.}
\label{fig_nullenB}
\end{figure*}

\begin{figure*}
\begin{tabular}{@{}lr@{}}
{\mbox{\includegraphics[scale=0.68,angle=0.]{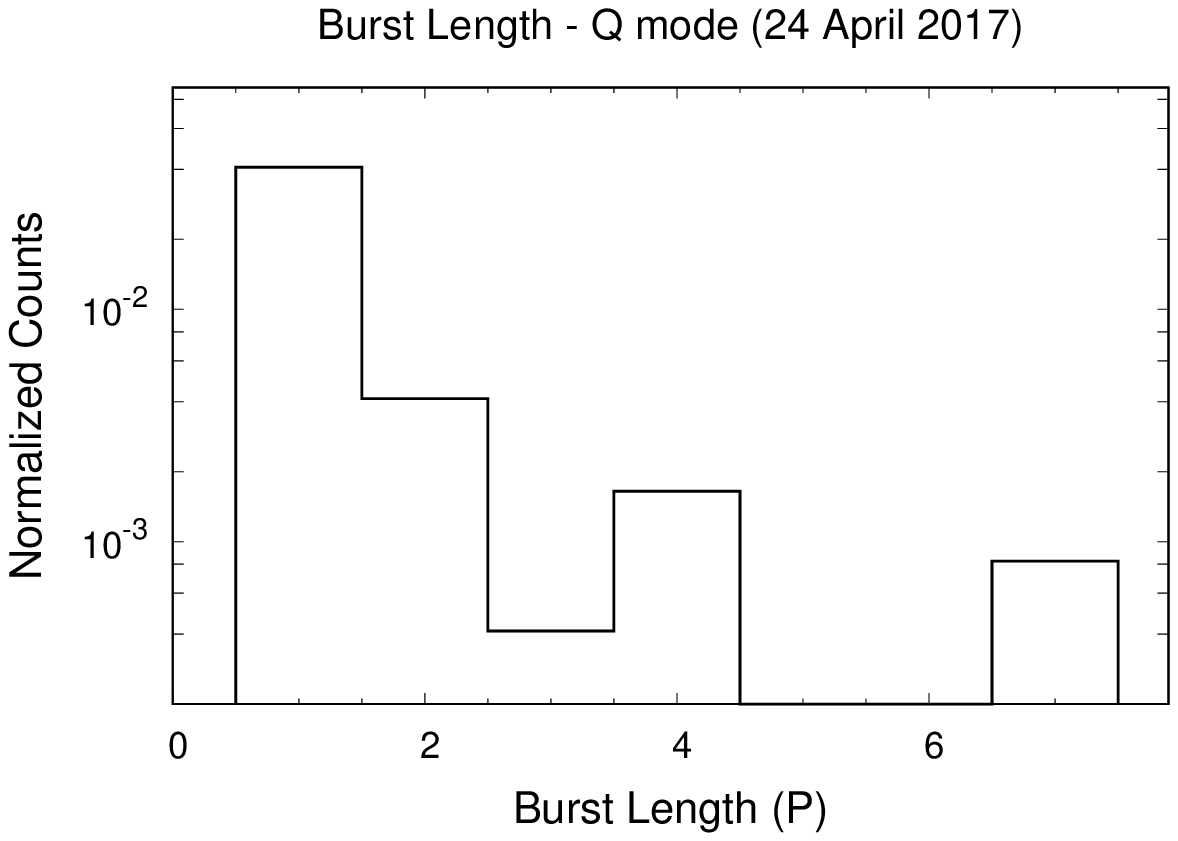}}} &
{\mbox{\includegraphics[scale=0.68,angle=0.]{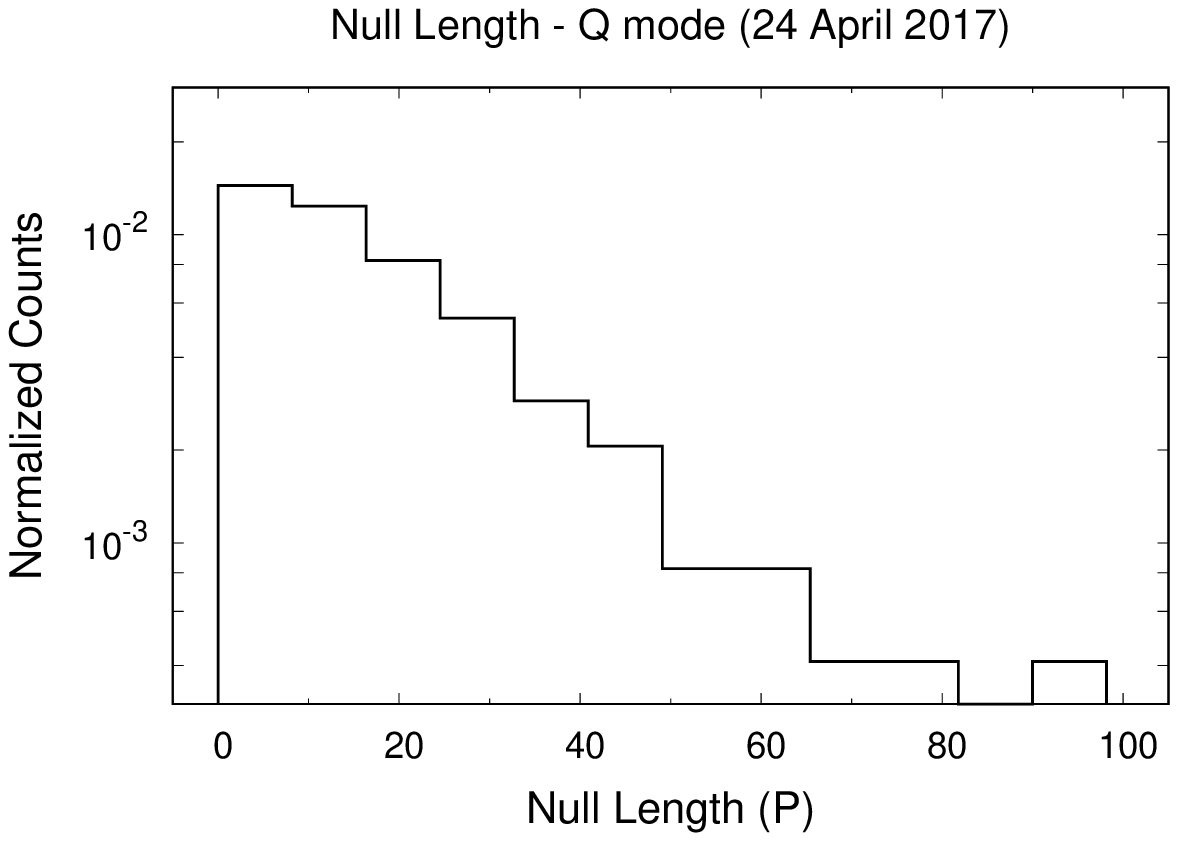}}} \\
\end{tabular}
\caption{The burst length (left panel) and null length (right panel) histograms
during B-mode, observed on 24 April, 2017. The pulsar shows short duration 
bursts in this mode lasting a few periods. The nulls are of much longer 
durations lasting several tens of periods at a time.}
\label{fig_nullenQ}
\end{figure*}

\noindent
The nulling varies a lot between the different emission modes. The nulling was 
more prominent during 24 April, 2017 when the pulsar was primarily in the 
Q-mode. A long duration nulling state was also observed on this day, lasting 
roughly 5500 periods ($\sim$ 48.6 minutes), before the onset of the Q-Bright 
mode. The pulsar showed nulls lasting tens of periods in between bursts of 
emission in the Q-mode. The B-mode on the other hand showed short duration 
nulls typically lasting one or two periods. The long duration nulls preceding 
the Q-Bright mode has also been reported in \citet{sob15} with durations of 
roughly an hour, and appears to be a separate emission state. The folded 
profile (see figure \ref{fig_Qfold}) did not show the presence of any low level 
emission during the long duration null state. Figure \ref{fig_modesngl} (right 
panel) shows the presence of nulling during the Q-mode, while figure 
\ref{fig_Bnull} shows examples of short nulls during B-mode. We have carried 
out a detailed analysis of nulling using the techniques described in 
\citet{bas17,bas18b,bas19b}. This included estimating the energy distributions 
for the on and off-pulse windows to determine the nulling fractions as well as
the distribution for consecutive null and burst pulses \citep{rit76}. An
example of the energy distributions during B-mode is shown in figure 
\ref{fig_Bdist}, where the main pulse and postcursor distributions are shown 
along with the off-pulse window. The nulling analysis was carried out for the 
main pulse since both the postcursor and the interpulse were not detected 
consistently at the single pulse level.

\begin{table}
\resizebox{\hsize}{!}{
\begin{minipage}{80mm}
\caption{Nulling properties during emission modes}
\centering
\begin{tabular}{ccccc}
\hline
 Date & Mode & $\langle BL\rangle$ & $\langle NL\rangle$ & NF \\
  &  & ($P$) & ($P$) & (\%) \\
\hline
  &  &  &  &  \\
 20 April & B & 28.5 & 1.12 & 3.8 \\
  &  &  &  &  \\
 22 April & B & 31.8 & 1.11 & 3.4 \\
  &  &  &  &  \\
 24 April & Q-Bright & 35.2 & 1.00 & 2.8 \\
  & Q & 1.69 & 19.0 & 91.9 \\
  &  &  &  &  \\
 26 April & B & 26.7 & 1.12 & 4.0 \\
  &  &  &  &  \\
 28 April & B & 24.8 & 1.14 & 4.4 \\
  &  &  &  &  \\
 30 April & B & 29.2 & 1.14 & 3.7 \\
  &  &  &  &  \\
\hline
\end{tabular}
\label{tabnull}
\end{minipage}
}
\end{table}

Table \ref{tabnull} lists the nulling behaviour during the different observing 
sessions, the average duration of the burst ($\langle BL\rangle$) and null 
($\langle NL\rangle$) states and the nulling fractions (NF) during each mode. 
The pulsar was in B-mode on five out of the six observing sessions and the 
nulling usually lasted for short durations typically one or two periods. The
burst states were usually of longer duration lasting several tens of periods 
which extended to a few hundred pulses on rare instances. In figure 
\ref{fig_nullenB} we show one example of the null length and burst length 
histograms during the B-mode, for the observations on 26 April, 2017. The
nulling behaviour was largely consistent for the different observing sessions 
in the B-mode. However, small variations were seen with the average burst 
lengths being shorter on 26 and 28 April, 26.7 $P$ and 24.8 $P$ respectively, 
and relatively longer, about 31.8 $P$, on 22 April. Similarly, the nulling 
fractions were also higher on 26 and 28 April, 4.0\% and 4.4\% respectively, 
compared to 22 April with nulling fraction of 3.4\%. The short duration 
Q-Bright mode had similar nulling behaviour to the B-mode, with the nulling 
being less frequent. Only single period nulls were seen in this mode while the 
average burst length was 35.2 $P$, which was higher compared to the B-mode. The
nulling fraction was 2.8\% which was also comparatively lower. In figure 
\ref{fig_nullenQ} the nulling behaviour during Q-mode is shown, where the null 
and burst length histograms show contrasting behaviour to the B-mode. The 
bursts were usually of short durations lasting one or more periods, while the 
nulls lasted for tens of periods at a time. The average null length was 19.0 
$P$ and the burst length was 1.69 $P$ during the Q-mode. The nulling fraction 
was also more than 90\% during this mode. \citet{sob15} also estimated the
nulling behaviour for the different emission modes. They reported nulling 
fraction of 1.8$\pm$0.5\% which is smaller than our observations. However, our
results show that the nulling behaviour does not remain constant during B-mode 
but shows variations across different observing sessions. They also reported 
nulling fraction of 80$\pm$9\% during the Q-mode which is consistent with our 
estimates within measurement errors.

\begin{figure*}
\begin{tabular}{@{}lr@{}}
{\mbox{\includegraphics[scale=0.72,angle=0.]{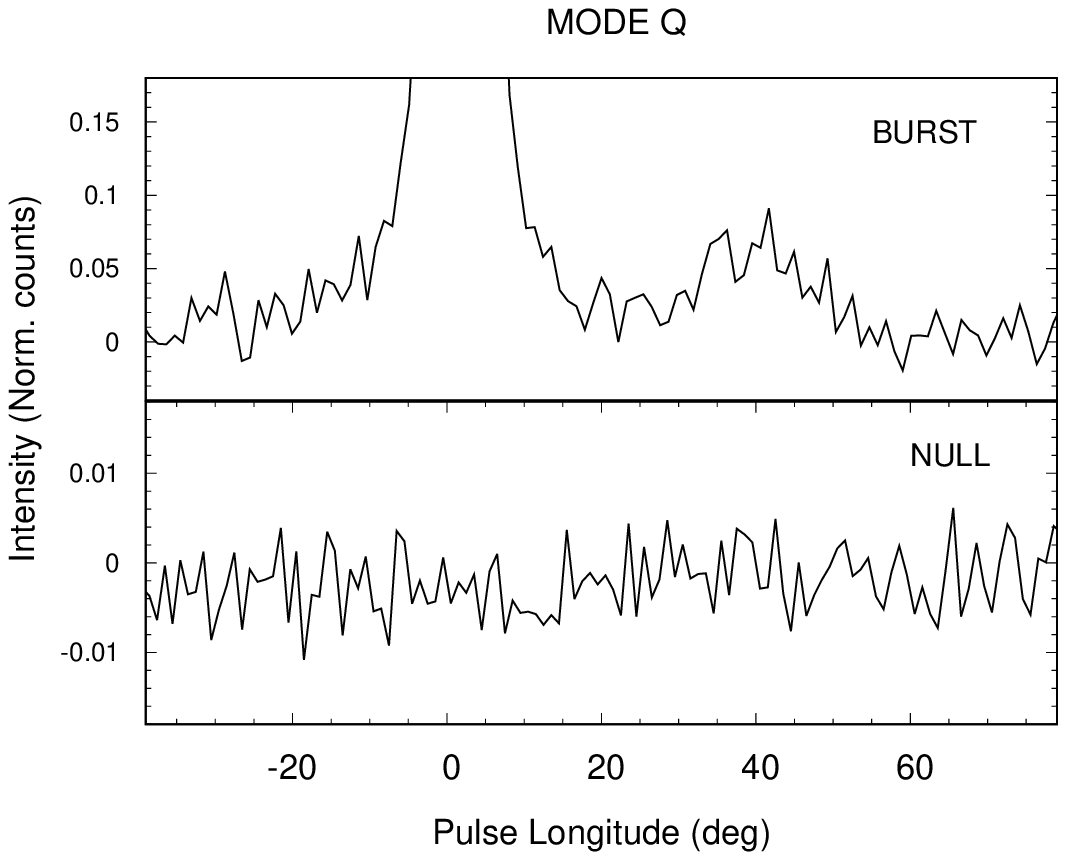}}} &
{\mbox{\includegraphics[scale=0.72,angle=0.]{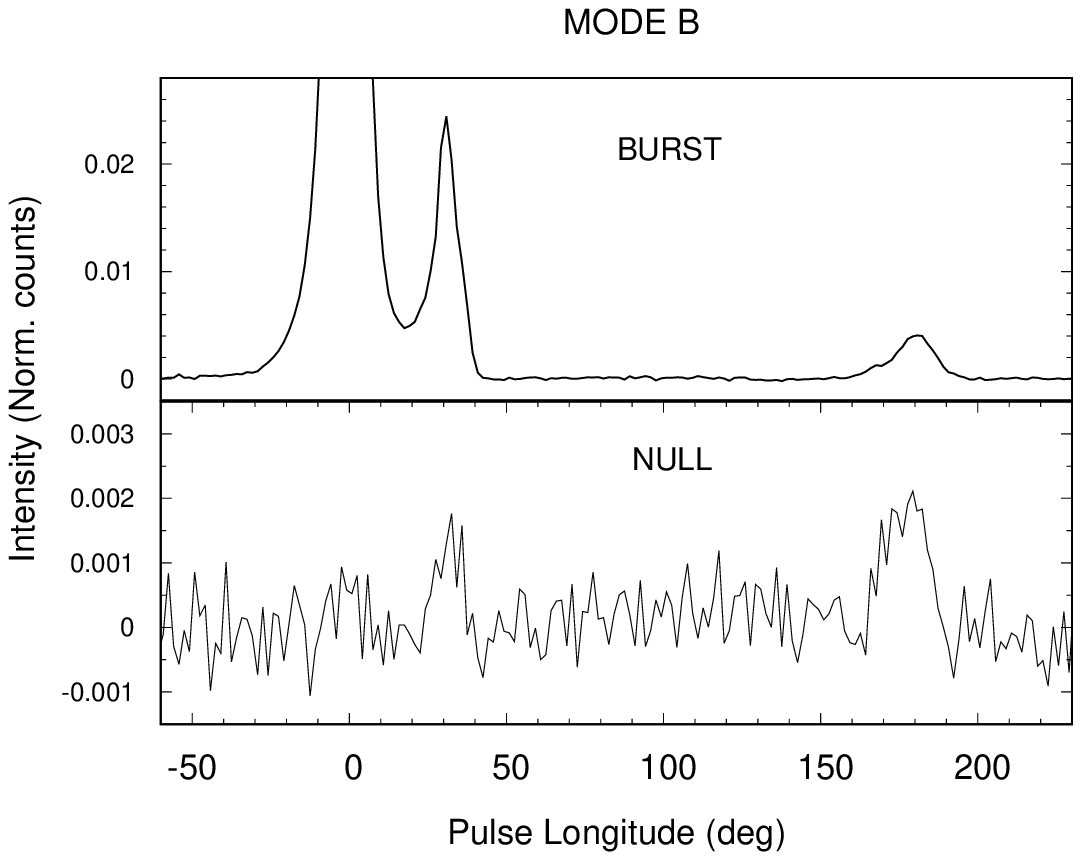}}} \\
\end{tabular}
\caption{The folded profile from the burst (top window) and null pulses 
(bottom window), for both Q-mode (left panel) and B-mode (right panel) are 
shown in the figure. The profiles during each mode have been scaled by the main
pulse peak of the burst profile. The folded profile from null pulses during the
Q-mode do not show the presence of any emission. On the other hand the 
corresponding profile in the B-mode shows low level emission in the postcursor 
as well as the interpulse. The postcursor component in the burst profile of the
Q-mode is clearly seen and more prominent than the average profile in the 
Q-mode.}
\label{fig_nullfold}
\end{figure*}

We have also estimated the folded profiles using just the null and burst pulses
in each of the modes to detect the presence of any low level emission during 
nulling. The average profiles in each case are shown in figure 
\ref{fig_nullfold}. In the Q-mode there was no emission seen during nulling for
any of the three components. The postcursor component was clearly seen in the 
burst profile of the Q-mode which was not detected in the average profile (see 
figure \ref{fig_Qfold}). The relative intensity of the postcursor component 
compared to the main pulse was more than 2-3 times higher in the Q-mode than in
B-mode. The null profile in the B-mode on the other hand showed the presence of
low level emission in the postcursor as well as the interpulse. However, no 
emission was seen in the main pulse during nulling as expected. The postcursor 
component in the null profile was much weaker and only around 5\% level of the 
burst profile. The interpulse on the other hand was comparable in the two 
profiles with the null profile showing more than 50\% of the burst profile 
intensity. 

\begin{figure*}
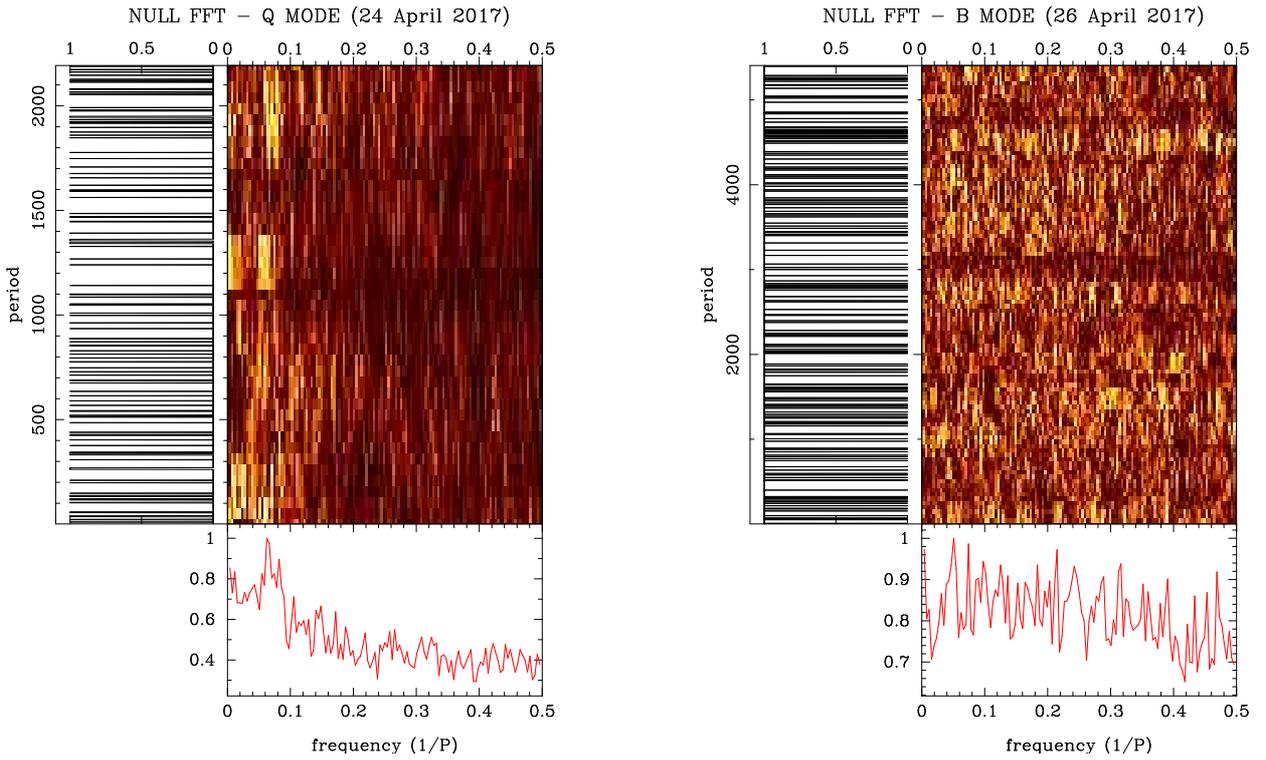

\begin{tabular}{@{}lr@{}}
{\mbox{\includegraphics[scale=0.38,angle=0.]{Qmode_nullfft.ps}}} &
\hspace{40px}
{\mbox{\includegraphics[scale=0.38,angle=0.]{Bmode_nullfft.ps}}} \\
\end{tabular}
\caption{The figure shows the time evolution of the nulling periodicity during
the Q-mode (left panel) B-mode (right panel). The nulls and burst periods were 
identified as `0' and `1' respectively, and Fourier transforms of the series 
were carried out to estimate the appearance of periodicity in the sequence. The
transitions from the null to the burst states during the Q-mode showed the 
appearance of wide low frequency structure indicating periodic behaviour for 
the longer duration nulls in this mode. The short nulls during the B-mode on 
the other hand did not exhibit any clear periodicity.}
\label{fig_nullfft}
\end{figure*}

We have also analysed the nulling behaviour during both the modes to 
investigate the appearance of periodicity during the transitions from null to 
burst states. The null and burst pulses were identified with `0' and `1' 
respectively, and Fourier transforms were carried out on these sequences as 
detailed in \citet{bas17}. In figure \ref{fig_nullfft} we show the results of 
this analysis on Q-mode (left panel) and B-mode (right panel). We report for 
the first time the appearance of periodic/quasi-periodic nulling in the 
Q-mode of the pulsar. No periodicity could be associated with transitions from 
the null to burst states in B-mode. In table \ref{tabampmod} we have estimated 
the properties of the periodic nulling feature seen in Q-mode. The nulling 
periodicity was estimated to be 14.2$\pm$3.3 $P$, which is higher than the 
periodic amplitude modulation in B-mode. The periodic nulling/amplitude 
modulation in many cases \citep{bas17,mit17} is not seen consistently over the 
entire duration of observation but becomes more prominent at certain sequences.
This is also seen for the Q-mode observations where in certain intervals the 
periodic behaviour almost vanishes.

\section{Periodic Modulation of Single Pulses}\label{sec:ampmod}

\begin{figure*}
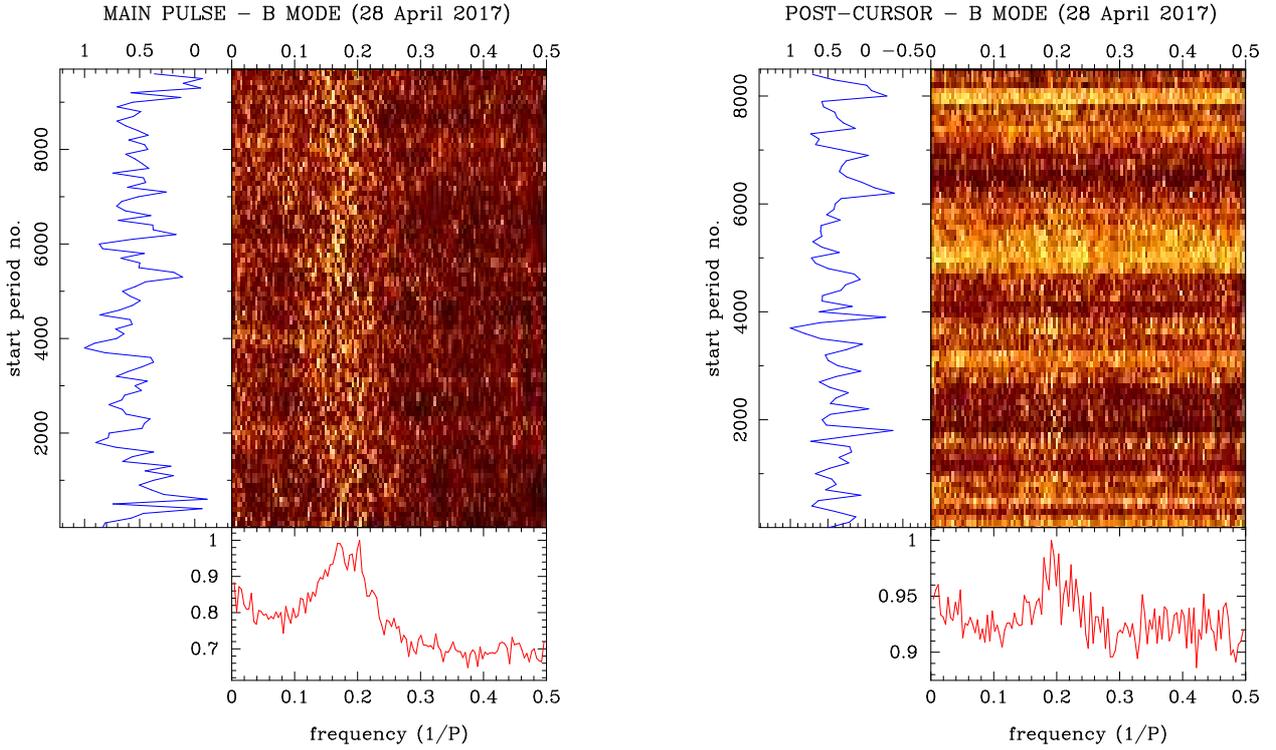

\begin{tabular}{@{}lr@{}}
{\mbox{\includegraphics[scale=0.38,angle=0.]{LRFSavg_MP.ps}}} &
\hspace{40px}
{\mbox{\includegraphics[scale=0.38,angle=0.]{LRFSavg_PPC.ps}}} \\
\end{tabular}
\caption{We have estimated the time evolution of the Longitude Resolved 
Fluctuation Spectra (LRFS) for the windows corresponding to the main pulse 
(left panel) and the postcursor (right panel) components. The time average 
fluctuation spectra shows a wide structure around 0.2 cycles/$P$ which becomes
apparent only after averaging over longer time intervals. The structure 
corresponds to periodic amplitude modulations in the intensity and is seen both
in the main pulse as well as the postcursor component.}
\label{fig_lrfsavg}
\end{figure*}

\begin{figure*}
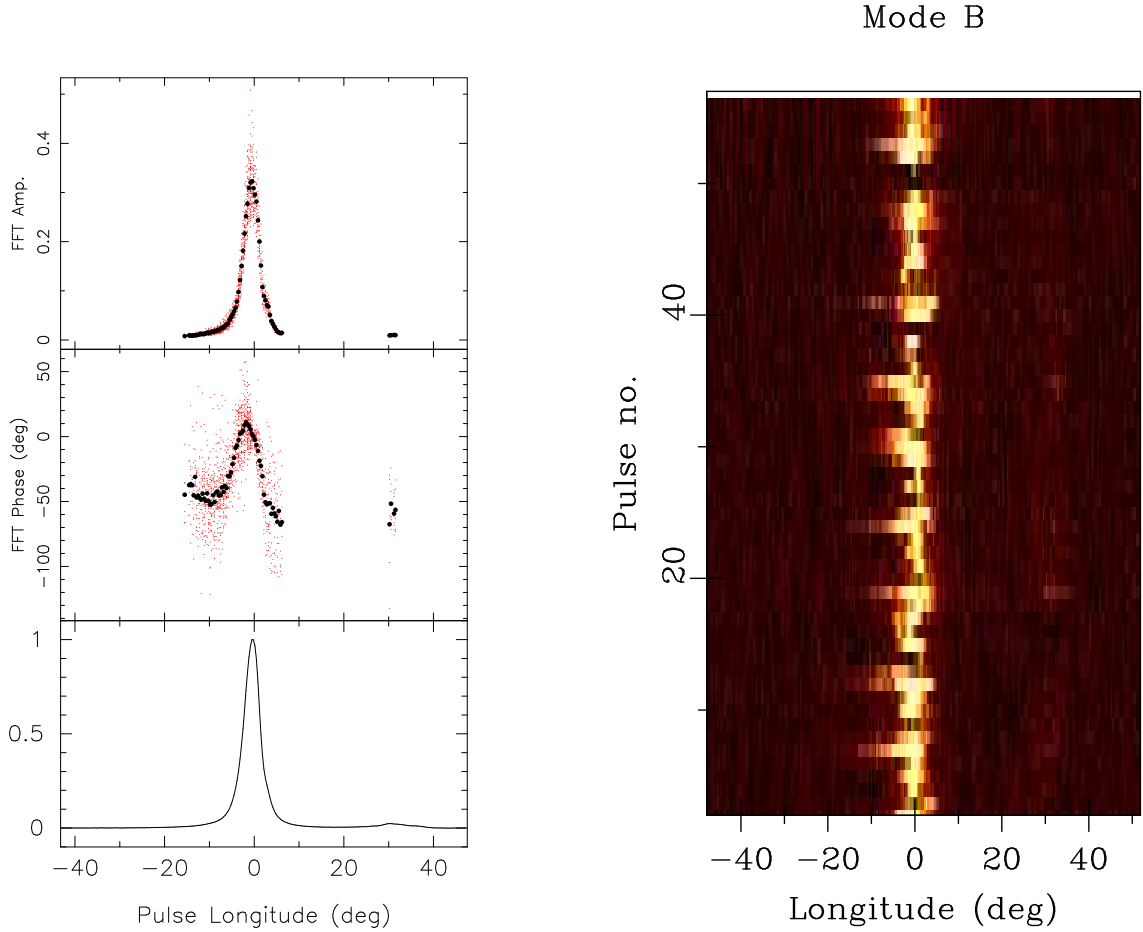

\begin{tabular}{@{}lr@{}}
{\mbox{\includegraphics[scale=0.45,angle=0.]{LRFSPhs.ps}}} &
\hspace{40px}
{\mbox{\includegraphics[scale=0.72,angle=0.]{J0826_Bampmod.ps}}} \\
\end{tabular}
\caption{The left panel of the figure shows the variation of the peak amplitude
(top window) of the fluctuation spectra in the B-mode corresponding to wide 
peak, $P_M \sim$ 5.5$P$, as well as the phase variations associated with 
the peak (middle window) across the pulse window. The phase variations in the 
main pulse resemble a bell shaped curve. The right panel shows a short section 
of the single pulse sequence in the B-mode responsible for the observed phase 
variations. The pulsar undergoes periodic amplitude modulations where the 
emission increases towards the edges of the pulse window in a periodic manner.}
\label{fig_phsvar}
\end{figure*}

\noindent
We have carried out a detailed analysis to study the periodic variations during
the different emission modes. We have used the fluctuation spectral analysis 
to estimate the periodicity in the single pulse behaviour as described in 
\citet{bas16}. The longitude resolved fluctuation spectra \citep[LRFS,][]{
bac73a} were estimated for sequences of 256 consecutive pulses at a time. 
Subsequently, the starting period was shifted by 100 pulses and the process was
repeated to generate a time evolution of the LRFS. In figure \ref{fig_lrfsavg} 
we show the average LRFS as a function of pulse period during the B-mode. We 
have separately estimated the spectra for the main pulse (left panel) as well 
as the postcursor component (right panel). The average spectra show the 
presence of a wide periodic structure seen in both the main pulse and the 
postcursor component. Table \ref{tabampmod} details the nature of the periodic 
structure, including the median peak frequency ($f_p$ = 0.18 cycles/$P$), Full 
width half maximum (FWHM = 0.08 cycles/$P$) of the peak, the $S$ factor defined
as the ratio between peak value and width, and the estimated periodicity ($P_A$
= 5.5$\pm$1.0 $P$). The error in the peak frequency ($\delta f_p$) was 
estimated as $\delta f_p$=FWHM/2$\sqrt{2\ln2}$ \citep{bas16}. The estimated 
periodicity during this mode was consistent with earlier measurements 
\citep{sob15}. No clear periodicity was seen in the average fluctuation spectra
during the Q-mode. The periodic phenomenon seen in the B-mode was a form of 
periodic amplitude modulation since no systematic drift band was seen in the 
single pulse sequence (see figure \ref{fig_phsvar}, right panel). The main 
pulse has been identified as a core component with the profile type being core 
single \citep{ran90}. The subpulse drifting is only seen in conal components of
pulsar profile with the effect absent in core \citep{ran86,bas19a}. This 
further indicates the phenomenon to be periodic amplitude modulation. We have 
also estimated phase variations across the pulse window corresponding to the 
peak frequency using the technique described in \citet{bas18a}. The phase 
behaviour is shown in figure \ref{fig_phsvar} (left panel, middle window) and 
represents a bell shaped curve across the main pulse. This behaviour was 
different from the periodic amplitude modulation cases studied in the past 
which usually show a constant phase value across the pulse window \citep{bas16,
bas19b}. The single pulse sequence shown in the right panel of the figure sheds
light into this unique phase behaviour. As the pulse intensity increases the 
pulse also becomes wider at both the leading and trailing edge of the main 
pulse, with the stronger pulses being much wider. This results in the phases 
showing an increasing slope in the leading edge and a decreasing slope towards
the trailing side. The increased intensities in the single pulses do not
necessarily mean that the peak amplitude is systematically changing. The 
increase in intensity can also arise from the emission originating from a wider 
region of the pulse window. This is contrary to subpulse drifting where the 
subpulse width does not change drastically between pulses but the subpulses 
usually shift from one edge of the window to the other resulting in systematic 
phase variations across the the entire component. The diffuse nature of the 
periodic behaviour indicates deviation from the usual behaviour, as seen in 
figure \ref{fig_Bnull}, with possibilities of narrow, bright pulses. There 
have been previously reported cases of phase reversals in the form of 
bi-drifting where the phases show opposite slopes \citep{bas18a}. However, the 
different phase behaviours are associated with different components and not 
within the same component as seen in this pulsar. \citet{wel06,wel07b} 
have carried out the 2-Dimensional Fluctuation spectrum \citep{edw02} for this 
pulsar, at 325 MHz and 1.4 GHz observing frequencies. They classified the 
pulsar as a diffuse drifter with a definite drift direction and large 
separation between adjacent drift bands, 70\degr~and 55\degr~respectively. The 
detailed phase behaviour reported in our study argues against the drifting 
interpretation reported in these earlier works. 

\begin{table}
\resizebox{\hsize}{!}{
\begin{minipage}{80mm}
\caption{Periodicity in the Emission Modes}
\centering
\begin{tabular}{c@{\hskip5pt}c@{\hskip5pt}c@{\hskip5pt}c@{\hskip5pt}c@{\hskip5pt}c}
\hline
 Type & MODE & $f_p$ & FWHM & S & P$_M$ \\
  &  & ($cy/P$) & ($cy/P$) & ($P/cy$) & ($P$) \\
\hline
  &  &  &  &  &  \\
Amp. Mod. & B & 0.182$\pm$0.035 & 0.082 & ~1.88 & ~5.51$\pm$1.05 \\
  &  &  &  &  &  \\
Periodic Null & Q & 0.070$\pm$0.016 & 0.038 & 15.96 & 14.2$\pm$3.3 \\
  &  &  &  &  &  \\
\hline
\end{tabular}
\label{tabampmod}
\end{minipage}
}
\end{table}

\section{High Intensity Emission in Single Pulses}\label{sec:post}

\begin{figure*}
\begin{tabular}{@{}lr@{}}
{\mbox{\includegraphics[scale=0.73,angle=0.]{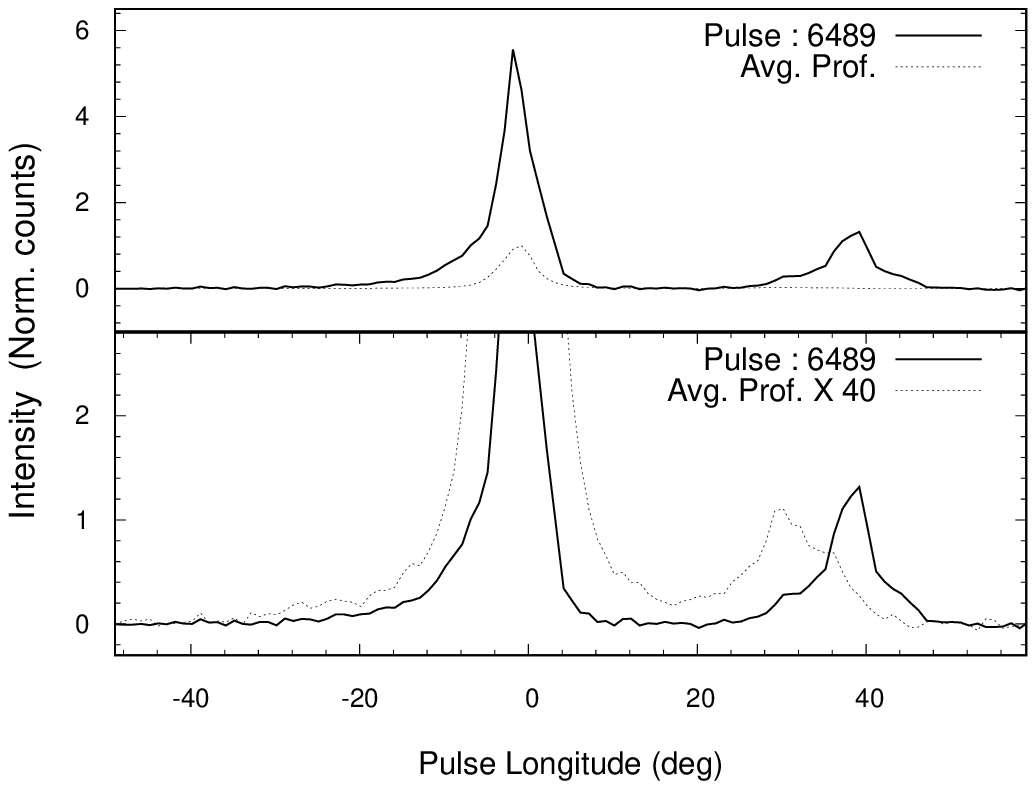}}} &
{\mbox{\includegraphics[scale=0.73,angle=0.]{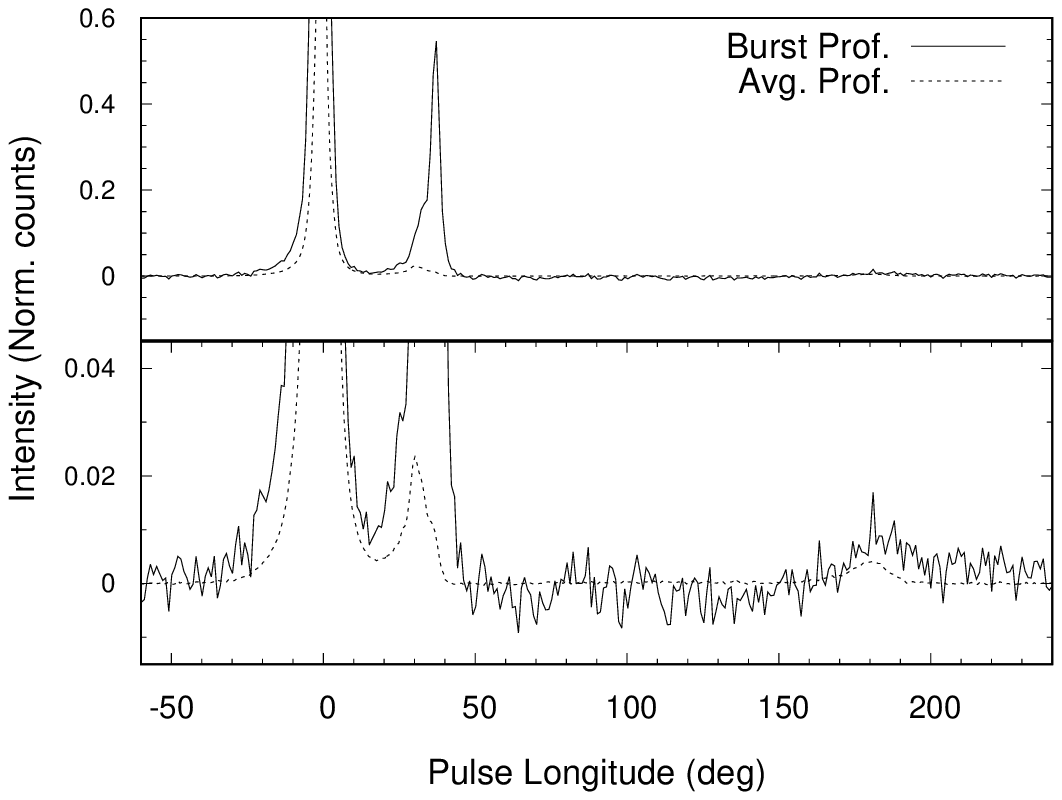}}} \\
\end{tabular}
\caption{The figure shows the nature of the high intensity emission seen in the
single pulses during rare instances. The left panel shows the comparison 
between the first pulse at the start of Q-Bright mode, pulse number 6489 from 
start of observation, and the average profile during this mode. The upper 
window presents the relative comparison between the two, where the main pulse 
peak is approximately 5 times the average profile peak (normalized to unity) 
and post-cursor peak is roughly 50 times the average profile peak. The lower
window shows an expanded version of the profile multiplied by a factor of 40. 
The postcursor in the bright pulse, peaks near the trailing edge in comparison 
with the average profile, while the bridge emission between the main pulse and 
the postcursor is barely detectable. The right panel shows the average profile 
from all the high intensity pulses (around 100) along with the average profile 
in the B-mode, whose peak is normalized once again to unity. The postcursor 
component in the average profile is more prominent near the trailing edge where
the average profile component is weaker. The lower window also shows an 
expanded view of the interpulse region, which does not show a commensurate 
increase in intensity.}
\label{fig_brightpuls}
\end{figure*}

\noindent
One of the features of the mode changing event from the null state to the 
Q-Bright state was the significant increase in the intensities of the main 
pulse and the postcursor component at the start of this mode (see the intensity
variations in figure \ref{fig_modesngl}, left panel). We have examined the 
properties of the first pulse in the Q-Bright mode in greater detail. In figure
\ref{fig_brightpuls} (left panel) we compare the first pulse with the average 
profile, while the estimates of the relative intensities are shown in table 
\ref{tabburst}. The main pulse intensity was more than 5 times compared to the 
average profile, but the postcursor component showed even more dramatic 
increase of almost 50 times the average. At this instance the postcursor peaked
near the trailing edge of the average component where negligible emission was 
seen otherwise. The bridge emission between the main pulse and the postcursor 
was barely detectable in this pulse. There was also no detectable emission seen 
in the interpulse region at the start of the Q-Bright mode. The interpulse 
emission in the average profile is much weaker at these observing frequencies 
and is not detected at the single pulse level. However, if the interpulse 
increased in an equivalent manner to the postcursor component then it would 
have likely been detected.

\begin{table}
\resizebox{\hsize}{!}{
\begin{minipage}{80mm}
\caption{Comparing first pulse during Q-Bright mode with average profile}
\centering
\begin{tabular}{c@{\hskip5pt}c@{\hskip5pt}c@{\hskip5pt}ccc}
\hline
$\frac{\textrm{MP}_{Bur}}{\textrm{MP}_{Avg}}$ & $\frac{\textrm{PC}_{Bur}}{\textrm{PC}_{Avg}}$ & $\frac{\textrm{MP}_{Avg}}{\textrm{PC}_{Avg}}$ & $\frac{\textrm{MP}_{Bur}}{\textrm{PC}_{Bur}}$ & \multicolumn{2}{c}{PC$_{Bur}$} \\
  &  &  &  & W$_{50}$ (\degr) & W$_{10}$ (\degr) \\
\hline
  &  &  &  &  &  \\
 5.56 & 47.60 & 36.1 & 4.2 & 5.2$\pm$0.3 & 17.7$\pm$0.3 \\
  &  &  &  &  &  \\
\hline
\end{tabular}
\label{tabburst}
\end{minipage}
}
\end{table}

\begin{table}
\resizebox{\hsize}{!}{
\begin{minipage}{80mm}
\caption{Statistics of Bursts in Post-cursor and Inter-pulse}
\centering
\begin{tabular}{c@{\hskip3pt}c@{\hskip3pt}c@{\hskip3pt}ccc@{\hskip2pt}c}
\hline
 Date & Mode &\multicolumn{3}{c}{NBursts} & \multicolumn{2}{c}{Burst Rate}\\
  &  & MP+PC & PC & IP & PC & IP \\
  &  &  &  &  & (/1000$P$) & (/1000$P$) \\
\hline
  &  &  &  &  &  &  \\
 20 April & B & 23 & 189 & $>$3 & 4.56 & --- \\
  &  &  &  &  &  &  \\
 22 April & B & 27 & 144 & --- & 4.49 & --- \\
  &  &  &  &  &  &  \\
 24 April & Q-Bright & ~2 & ~~3 & --- & 5.91 & --- \\
  &  &  &  &  &  &  \\
 26 April & B & 14 & ~96 & 20 & 3.69 & 0.77 \\
  &  &  &  &  &  &  \\
 28 April & B & 20 & 151 & $>$6 & 4.93 & --- \\
  &  &  &  &  &  &  \\
 30 April & B & 23 & 104 & $>$6 & 4.21 & --- \\
  &  &  &  &  &  &  \\
\hline
\end{tabular}
\label{tabburststat}
\end{minipage}
}
\end{table}

The lack of increased intensity in the interpulse component might suggest the 
mode changing from the null state to the Q-Bright state to be restricted to a 
single pole of the pulsar. Alternatively, the increased intensities of the main
pulse and particularly the postcursor component at the start of the mode could 
also be merely coincidental. Other observations during the transition to the 
Q-Bright state are necessary to validate whether such transitions are always 
associated with increased intensities. However, in the absence of such 
observations we have investigated in detail the occurrence of increased 
intensities of the different components in the single pulse behaviour of the 
pulsar particularly in the B-mode. The pulse energy distributions of the 
different components have been reported in previous studies \citep{sob15}, 
where the main pulse and the postcursor exhibited log-normal distributions. Our 
estimates of the distributions were also consistent with the earlier studies. 
The main pulse had maximum intensities around 10-15 times the average 
intensity. However, more dramatic increases in intensities were seen in the 
postcursor component. In addition to the first pulse in the Q-Bright mode, 
which had a peak intensity of 47.60 times the average intensity, there were 
several instances where the postcursor intensity exceeded 100 times the average
value, an order of magnitude greater than the main pulse high intensity 
emission. In these rare bright cases the postcursor peaked near the trailing 
edge of the average component. We have listed in Table \ref{tabburststat} the 
number of pulses during each observing run where the postcursor intensity 
exceeded 20 times the average value, as well as the rate of such events 
every 1000 periods. As expected the high intensity in the postcursor were 
extremely rare with roughly 4-5 such instances seen every 1000 periods. The 
Table also lists the number of pulses where both the postcursor and the main 
pulse have increased intensity, i.e the postcursor with more than 20 times its 
average value and the main pulse with more than 5 times its average intensity. 
Such cases accounted for 15-20\% of the total instances of bright postcursor 
events. We have estimated the average profile from the pulses with these bright
postcursor components as shown in figure \ref{fig_brightpuls} (right panel). 
The postcursor in this average profile was stronger near the trailing edge 
similar to the first pulse in the Q-Bright mode. The main pulse was also 
stronger than the average suggesting that the overall intensity increases 
during these events. However, the interpulse did not show any discernible 
increase in intensity indicating that these bright emission states were 
restricted to a single pole.

\begin{figure}
\begin{center}
\includegraphics[scale=0.68,angle=0.]{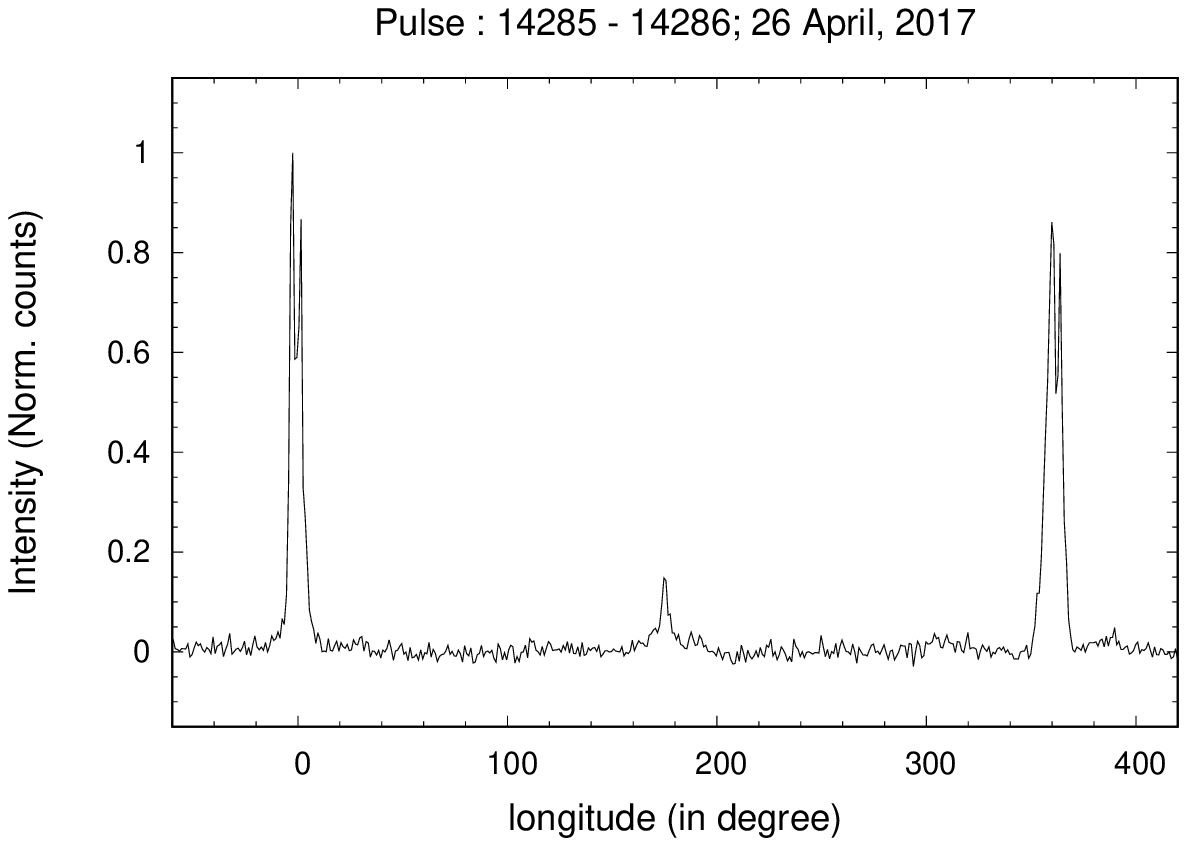}
\end{center}
\caption{The figure shows the prominent interpulse emission observed between 
pulse 14285 and 14286 on 26 April, 2017. This is a rare event where the 
interpulse is more than 40 times the average intensity. The intensities in the 
plot is normalized by the average profile peak.}
\label{fig_intburst}
\end{figure}

We have also investigated the interpulse emission in the single pulses to check
the possibility of increased emission. This was particularly challenging due to
the interpulse being intrinsically very weak and not visible at the single 
pulse level. The single pulse emission with 20 times the average interpulse 
intensity was just at the detection limit. We report the presence of such 
bright interpulse emission in the B-mode of this pulsar. An example is shown in
figure \ref{fig_intburst} between pulse 14285 and 14286, observed on 26 April. 
The interpulse in this instance was around 15\% of the average main pulse 
intensity and more than 40 times the average interpulse intensity. The baseline
level was particularly stable on 26 April with few systematics affecting the 
observations and was most conducive for these studies. We observed around 20 
cases of interpulse emission which were more than 20 times the average 
intensity and hence detectable. This suggested such high intensity emission 
states in the interpulse to be extremely rare with less than 1 such event every
1000 periods. On the other observing days the noise at the baseline levels were
higher and we could only detect a few of the stronger intensity events from the 
interpulse region. We did not detect any significantly increased emission in 
either the main pulse or the postcursor component for any of the instances 
where the interpulse was very bright. We have estimated the peak 
intensities of the two main pulses preceding and following the 20 very bright 
interpulse events on 26 April. The preceding main pulses had minimum intensity
of 0.5 times the average profile peak, maximum peak of 4.0 times the average 
peak with mean ratio being 1.7. In the case of the main pulses following these 
events the minimum peak was 0.1 times the average profile peak, the maximum 
ratio was 1.7 times and the mean value being 0.8. This indicates that the 
preceding main pulse corresponding to the bright interpulse events might be 
more energetic than the pulses following these events. However, this dependence
cannot be claimed with great confidence due to the small number of such high 
intensity interpulse events observed. There are also no clear indication of the 
main pulse energies on either side of these bright interpulses to be along any 
narrow region of the main pulse energy distribution.

\section{Discussion}\label{sec:disc}

\subsection{The periodic modulations in the different Emission Modes}
\noindent
The periodic variations in the single pulse sequences of pulsars have many 
different forms with likely varying underlying physical mechanisms responsible 
for them. The most well known is the phenomenon of subpulse drifting where one 
or more subpulses within the pulse window shows systematic drift bands. The 
subpulse drifting is believed to originate due to {\bf E}x{\bf B} drift of the
sparking discharges in the inner acceleration region \citep[IAR,][]{rud75}. The 
sparks subsequently generate the plasma responsible for the radio emission 
where the drifting behaviour is imprinted \citep{mel00,lak19}. In addition to 
subpulse drifting certain pulsars also show the presence of periodic amplitude 
modulations and periodic nulling \citep{bas16,bas17}. In these cases the pulse 
intensities show periodic/quasi-periodic variations in intensity without any 
systematic subpulse motion within the pulse window. It is believed that the 
periodic amplitude modulation/nulling is not related to the {\bf E}x{\bf B} 
drift of the sparks but rather due to a triggering mechanism periodically 
modifying the pair production process in the pulsar magnetosphere 
\citep{bas17}. It has also been found in recent works that there are clear 
physical differences seen in subpulse drifting and periodic amplitude 
modulation/nulling. Subpulse drifting is only seen in the conal components of 
pulsars, while the central core component does not exhibit drifting. Periodic 
amplitude modulation as well as periodic nulling encompasses the entire pulse 
window \citep{bas19a}. Additionally, subpulse drifting shows subpulse motion or
shifts within the pulse window. In fluctuation spectral analyses \citep{bac73a,
bas16} the subpulse motion within the pulse window is reflected as systematic 
phase variations corresponding to the peak frequency. Periodic amplitude 
modulation and periodic nulling are usually phase stationary since they affect 
the entire pulse window. There are also clear differences in the physical 
properties of pulsars showing these different phenomena. Drifting is primarily 
seen in pulsars with spin down energy loss ($\dot{E}$) $<$ 5$\times$10$^{32}$
erg~s$^{-1}$ \citep{bas16,bas19a}. On the other hand pulsars with periodic 
amplitude modulation/periodic nulling do not show any such cutoff in their 
$\dot{E}$ values. There is also a correlation seen between the drifting 
periodicities with $\dot{E}$, such that pulsars with lower $\dot{E}$ tend to 
have higher drifting periodicities. Such dependence of periodicity on the 
$\dot{E}$ is absent for the periodic amplitude modulations and nulling.

The periodic amplitude modulations and nulling seen in the different emission 
modes of the pulsar J0826+2637 exhibit a consistent behaviour. The pulsar has
$\dot{E}$ = 4.5$\times$10$^{32}$~erg~s$^{-1}$ which is at the boundary for 
drifting pulsars and its main pulse corresponds to a core component. However,
the periodic behaviour still provides a lot of variations from typical 
examples. As discussed earlier the phase variations corresponding to the 
amplitude modulations in B-mode are not flat but show phase changes at both the
leading and trailing edges of the main pulse. This is one of the first examples
where we see not only a periodic change in intensity but also single pulse 
width. Incidentally a U-shaped phase behaviour is also seen for the periodic 
modulations of the main pulse in the pulsar J1705$-$1906 (B1702$-$19), another 
pulsar with interpulse emission \citep{wel07a}. The modulations are most 
prominent in the trailing edge and seems to be a form of periodic amplitude 
modulations. However, it is not clear if an equivalent change in the single 
pulse widths can be associated with them. The periodicity of the modulations in
the B-mode of PSR J0826+2637 (5.5$P$) is higher than expected from subpulse 
drifting ($\leq$2$P$ at the $\dot{E}$ boundary), but are still at the lower end
for typical amplitude modulation and periodic nulling cases which have 
periodicities between 10-100$P$. In addition to the main pulse the postcursor 
component also exhibits periodic amplitude modulations. This is one of the few 
instances where such periodic behaviour has been associated with either 
pre/postcursor (PPC) features. As discussed in \cite{bas15} the PPC emission 
originates outside the conventional emission region of the main pulse and hence
is expected to have a different emission mechanism. The presence of periodic 
amplitude modulations in the postcursor further reiterates the proposition in 
\cite{bas17} that the underlying physical processes are related to plasma 
generation in the IAR and independent of the radio emission mechanism. On the 
other hand the periodic nulling seen in the Q-mode has higher periodicities 
(14.2$P$) compared to the modulations in the B-mode, and is a more typical 
example of this phenomenon. The difference in the periodicities suggests that 
during mode changing the underlying triggering mechanism also undergoes 
changes. A number of pulsars have been studied where the periodic amplitude 
modulations and periodic nulling seem to show temporal changes in their 
periodic properties. As discussed in more detail in the next subsection, the 
two modes of PSR J1825$-$0935 (B1822$-$09) show the presence of periodic 
modulations with different periodicities \citep{lat12}. The core/cone Triple 
pulsar J1948+3540 (B1946+35) shows the presence of wide and diffuse periodic 
amplitude modulations in the average fluctuation spectra \citep{mit16}. 
However, there are instances when very sharp features resembling highly ordered
periodic behaviours are seen. However, no associated mode changing was reported
for this source. The presence of periodic nulling was also reported for the 
pulsar J2006$-$0807 (B2003$-$08) with a Multiple profile comprising of a 
central core and two pairs of conal components \citep{bas19b}. The pulsar 
exhibits the presence of four clear emission modes two of which, A-mode and 
B-mode, show the presence of subpulse drifting. The periodic nulling is only 
seen in C-mode and D-mode. The pulsar J0826+2637 provides another important 
addition to this newly emergent phenomenon in pulsars.

\subsection{The relation between the radio emission from the two poles}
\noindent
The relationship between the radio emission from the main pulse and interpulse
has important implications for understanding the physical processes in pulsars.
There have been reports of interaction between the main pulse and interpulse 
particularly in three pulsars J1057$-$5226 (B1055$-$52) \citep{big90,wel09,
wel12}, J1705$-$1906 \citep{wel07a} and J1825$-$0935 \citep{fow82,gil94,bac10,
lat12,her17,yan19}. In case of PSR J1057$-$5226 both the main pulse and the 
interpulse show prominent profiles with the main pulse having four components 
and the interpulse with three components in the average profile. Both the main 
pulse and the interpulse show roughly phase stationary, periodic amplitude 
modulations of $\sim$20$P$. Additionally, the phases between the main pulse and
the interpulse are offset by $\sim$40\degr~which remains constant over time. 
Periodic amplitude modulations are also observed for pulsar J1705$-$1906 where 
the main pulse as well as the interpulse show periodic variations of 
$\sim$10$P$, and they appear to be phase locked over observations spanning 
several years. PSR J1825$-$0935 has a precursor component preceding the main 
pulse in addition to the interpulse emission. The pulsar exhibits two modes, 
B-mode and Q-mode similar to PSR J0826+2637 and shows periodic amplitude 
modulations of $\sim$40$P$ in the Q-mode for both the interpulse and the main 
pulse component which also seem to be phase locked. The precursor emission is 
absent during the Q-mode. On the contrary the interpulse vanishes during the 
B-mode when the precursor is most prominent. The B-mode also shows periodic 
amplitude modulation with a different periodicity $\sim$70$P$. The emission 
modes are short lived lasting several hundred periods at a time. \cite{lat12} 
have also reported instances of mode mixing when both the precursor and 
interpulse exist simultaneously. These observations suggest that there is a 
possibility of interaction between the two poles of the pulsar magnetosphere in
a manner that periodically affects the radio emission. All three pulsars seem 
to exhibit periodic amplitude modulations and not subpulse drifting. 

In the case of PSR J0826+2637 it was not possible to detect any modulation 
features for its weak interpulse emission. But the modal behaviours still 
provided interesting insights regarding the interaction between the two poles. 
Firstly, no interpulse emission was seen during the null state preceding the 
Q-bright mode on 24 April, 2017, suggesting that the radio emission was absent 
or greatly reduced beyond detection limits for both poles. Additionally, during
the Q-mode when the main pulse and postcursor intensity decreased 
significantly, the interpulse was not detected in our observations. However, 
there was also differing behaviour between the main pulse and the interpulse 
emission particularly during the B-mode. Both the postcursor component and the 
interpulse showed higher intensity emission in the single pulses which was at 
least 20 times greater than the average value. The only instance of transition 
from the null state to the Q-bright mode showed increased intensity from both 
the main pulse and the postcursor component but no such variation was seen in 
the interpulse. There was no clear evidence of any association between the high
intensity emission from the two poles during B-mode. Additionally, during the 
short duration nulls of the main pulse in this mode the interpulse emission did
not seem to change. On the other hand, during nulls in the Q-mode no emission 
was seen in the interpulse as well. The longer nulls in Q-mode were periodic in
nature while the short duration nulls in the B-mode did not show any 
periodicity. There are indications that the emission changes are affected 
simultaneously in the main pulse and interpulse. However, the interpulse 
emission in most of our observations is too weak to draw any definitive 
conclusions regarding the interaction between them and more sensitive 
observations are required to answer this question.

The different emission features seen in pulsar J0826+2637 indicate the presence
of diverse physical mechanisms within the pulsar magnetosphere. There is 
increasing observational evidence that the changes in the radio emission are 
driven by variations seen at different scales within the pulsar magnetosphere. 
Phenomena such as subpulse drifting, micro-structures in the single pulses, 
short duration nulls not associated with periodicity, etc., are likely affected
by local variations related to the plasma generation and the subsequent radio
emission mechanism \citep{lak19}. These variations generally affect a few 
pulses at a time. On the other hand a separate class of emission properties 
have emerged which seem to be pan-magnetospheric. These include the 
intermittent behaviour in pulsars which show changes in period derivatives 
during long nulling intervals lasting from weeks to months \citep{kra06,lyn17},
the transition of the pulsar between different emission modes, including long 
duration null modes, with modal durations from several hundred periods to 
several hours at a time, the periodic intensity modulations with durations of 
10-100 periods, amongst others. The pan-magnetospheric nature of the later two 
phenomena, mode changing and periodic intensity modulation, is highlighted by 
the interaction between the main pulse and the interpulse which appear to be 
simultaneously affected during these changes. The radio emission shows 
different emission behaviour during mode changing, like changes in emission 
intensity, the nature of subpulse drifting, etc. However, the radio emission 
heights and the corresponding location of the emission region remains largely 
unchanged during the different modes \citep{bas18b,bas19b}.

\subsection{The X-ray variation during the different emission modes}
\noindent
Pulsar J0826+2637 was observed simultaneously in the 0.2-2 keV X-ray band 
\citep{her18}. The pulsar exhibited synchronous mode changing in radio and 
X-ray frequencies, with the X-ray emission showing significant detection during
the B-mode and Q-bright mode and non-detection during the null and Q-mode. The 
X-ray observations were separated into one hour durations and showed up to 20\%
variability during the B-mode. It was speculated that such variations were 
closely connected with the dynamics of the radio emission. The lack of absolute 
flux calibration during the radio observations make it difficult to verify 
whether an equivalent change in intensity was present in the radio emission. 
However, we investigated correlations between other aspects of the emission 
dynamics during the B-mode. The B-mode showed short duration nulls which were 
present in less than 5\% of the total pulses. The X-ray intensities on 20 and 
22 April had instances of 15-20\% higher intensities compared to the 
observations on the remaining days. The nulling fractions also showed slightly 
lower values on these days. We have estimated the nulling fractions 
corresponding to the one hour windows over which the X-ray variability were 
calculated. The nulling fractions varied between 3-5\% within each observing 
session as well as across all the sessions.
However, no discernible trend with the X-ray variability was seen. 
Additionally, there was no clear difference in the rate of high intensity 
postcursor emission during the different days when the pulsar was observed in 
the B-mode. We conclude that given the low counting statistics of the X-ray 
observations ($\leq$ 0.01 counts/sec), no clear evidence of correlated 
short-term variability of the radio and X-ray emission during B-mode could be 
detected.


The synchronous X-ray and radio mode switching from PSR J0826+2637 makes it 
only the second pulsar after J0946+0951 (B0943+10) \citep{her13} where such 
behaviour could be detected. Both these sources exhibit two distinct modes 
characterized by their `Quiet' and `Bright' emission states which can last for 
several hours at a time \citep{bac11,sob15}. The X-ray emission on the other 
hand has contrasting behaviour for the two pulsars. The X-ray was not 
detectable in the Q-mode of PSR J0826+2637 which was at least $\sim$9 times 
less than the detected flux during the B-mode. While in the case of PSR 
J0946+0951 the X-ray emission showed an increment of $\sim$2.4 times in the 
Q-mode compared to the B-mode, with the X-ray pulsed fraction changing as a 
function of energy \citep{mer16}. Additionally, simultaneous radio and X-ray 
observations were also conducted for PSR J1825$-$0935 which exhibits shorter 
duration modes lasting several hundred periods \citep{her17}. However, no 
variation in the X-ray emission during mode changing was detected for this 
source. Similar studies were carried out for the pulsar J0614+2229 (B0611+22) 
which show presence of quasi-periodic bursting emission \citep{raj16}. The 
pulsar was observed simultaneously with \textit{XMM-Newton}, but no detectable 
X-ray emission could be seen. At present our understanding of the physical 
processes in the pulsar magnetosphere is limited and hence any possible 
explanations for the simultaneous radio and X-ray mode changing is mostly 
speculative\footnote{\cite{her18} have discussed a number of possible scenarios
leading to simultaneous radio and X-ray mode changing.}. Though it seems more 
likely that the X-ray variations reflect the pan-magnetospheric changes 
discussed earlier. There does not appear to be a short duration association 
between the radio and X-ray emission, however, more detailed studies with 
(future) X-ray missions with larger sensitive areas as well as flux calibrated 
radio observations are necessary to further explore any such correlation. There
is one aspect of the physical parameters that have not gathered much attention 
but seems interesting and relevant for this phenomenon. There is increasing 
evidence that a phase transition in the emission properties of pulsars is seen 
between $\dot{E} \sim$ 10$^{32}$-10$^{33}$~erg~s$^{-1}$. As discussed earlier 
the subpulse drifting is only seen in pulsars with $\dot{E}$ below this range. 
A change in the average polarization behaviour has also been reported 
associated with $\dot{E}$ \citep{wel08,mit16}. The percentage polarization in 
pulsars is high usually exceeding 60\% in the high $\dot{E}$ range, but 
decreases near the transition region to less than 20\% and subsequently 
increases slightly to around 30\% for lower $\dot{E}$ pulsars. Additionally, 
the five pulsars which show intermittent behaviour, i.e where the radio 
emission vanishes for long durations, all lie in this narrow $\dot{E}$ range 
\citep{lyn17}. Of the three pulsars J0826+2637, J0946+0951 and J1825-0935, 
where detailed simultaneous radio and X-ray studies have been carried out the 
first two showed synchronous variations. The three pulsars have $\dot{E}$ 
values of 4.52$\times$10$^{32}$ erg~s$^{-1}$, 1.04$\times$10$^{32}$ 
erg~s$^{-1}$ and 4.56$\times$10$^{33}$ erg~s$^{-1}$, respectively, indicating a
possible connection with this narrow boundary. However, it is clear that this 
association is mostly speculative without any detailed models explaining the 
origin of such transitions in the $\dot{E}$ parameter and the lack of 
simultaneous radio and X-ray observations from a significant number of pulsars.

\section{Summary and Conclusion}\label{sec:sum}
\noindent
We have carried out a detailed study of the single pulse behaviour from the 
different emission states of the pulsar J0826+2637. The pulsar exhibited two 
prominent modes, B-mode and Q-mode which lasted for hours at a time, in 
addition to a short lived null state as well as the Q-bright mode. The two 
primary modes were distinguished by the nature of nulling behaviour. In B-mode
the main pulse nulled for short durations lasting a few periods and nulling 
fraction less than 5\%. In contrast, the pulsar nulled for more than 90\% of 
duration in Q-mode which was interspersed with short bursts of emission. The 
nulling during B-mode was localised to the main pulse with the postcursor 
showing lower level emission and the interpulse emission was unchanged during 
the main pulse nulls. On the contrary no emission was seen in any of the 
components during the nulls in the Q-mode. We have verified the presence of 
periodic amplitude modulations of intensity ($\sim$5$P$) in the main pulse 
during B-mode, which was also seen to be present in the postcursor component. 
It was found that during these periodic modulations in addition to change in 
intensity the main pulse also showed systematic variation in widths. 
Additionally, we have also discovered the appearance of periodic nulling in the
Q-mode with a different periodicity ($\sim$14$P$). The pulsar joins a small 
group where different periodic amplitude modulations/periodic nulling are seen
to develop in their single pulse behaviour. There were instances where the 
postcursor and the interpulse components showed higher intensity emission which
were at least more than 20 times the average value. The most prominent amongst 
these was the first pulse during the transition from the null mode to the 
Q-bright state where the main pulse also had around five times the average 
intensity. However, more observations are required to verify if this was a 
coincidental effect or a by-product of the underlying physical processes 
responsible for such changes. The main pulse also had higher intensities during
the high intensity emission from the postcursor, but no connection was seen 
between increased emission states from the two poles. Apart from the large 
scale synchronous variations in the radio and X-ray emission during the Q-mode
and B-mode, we did not detect any clear indications of small scale (15-20\%) 
X-ray variations in the single pulse dynamics of the radio emission. There has 
been evidence for periodic amplitude modulations to be correlated in the two 
poles of the pulsar magnetosphere. Despite the lack of single pulse emission in
the interpulse, we have found indications of the periodic nulling in Q-mode to 
be present in both poles. The periodic amplitude modulations and periodic 
nulling are believed to be a result of a triggering mechanism periodically 
modifying the pair production process in the pulsar magnetosphere. These 
observations indicate that the triggering mechanism is pan-magnetospheric, 
affecting both the poles at the same time.

\section*{Acknowledgments}
We thank the referee for the comments which helped to improve the paper. We 
thank Prof. Wim Hermsen and Prof. Joanna Rankin for their comments. DM 
acknowledges funding from the grant ``Indo-French Centre for the Promotion of 
Advanced Research - CEFIPRA". We thank the staff of the GMRT who have made 
these observations possible. The GMRT is run by the National Centre for Radio 
Astrophysics of the Tata Institute of Fundamental Research.

\end{document}